\documentclass[10pt,twocolumn,letterpaper,pagenumbers]{article}

 \usepackage{iccv}              
\PassOptionsToPackage{numbers, compress, sort}{natbib}
\definecolor{iccvblue}{rgb}{0.21,0.49,0.74}
\usepackage[pagebackref,breaklinks,colorlinks,allcolors=iccvblue]{hyperref}

\usepackage{microtype}
\usepackage{graphicx}
\usepackage{booktabs}

\usepackage{wrapfig}
\usepackage{pifont}
\usepackage{color, colortbl}

\usepackage{blindtext}
\usepackage{lipsum}

\usepackage{multirow}
\usepackage{graphicx}
\usepackage{listings}

\usepackage{bbm}

\usepackage [english]{babel}
\usepackage [autostyle, english = american]{csquotes}

\usepackage{amsmath}
\usepackage{amssymb}
\usepackage{mathtools}
\usepackage{amsthm}

\usepackage[capitalize,noabbrev]{cleveref}

\theoremstyle{plain}

\theoremstyle{definition}

\theoremstyle{remark}

\usepackage{pifont}
\usepackage{url}
\usepackage[most]{tcolorbox}
\usepackage{lipsum}
\usepackage{wrapfig}
\usepackage{booktabs}
\usepackage{multirow,mathtools } 

\usepackage{adjustbox}
\MakeOuterQuote{"}

\usepackage{amsmath,amsfonts,bm}

\def\eqref#1{(\ref{#1})}

\def\1{\bm{1}}

\newcommand{\Dr}{\mathcal{D_{\mathrm{r}}}}
\newcommand{\Df}{\mathcal{D_{\mathrm{f}}}}

\DeclareMathAlphabet{\mathsfit}{\encodingdefault}{\sfdefault}{m}{sl}
\SetMathAlphabet{\mathsfit}{bold}{\encodingdefault}{\sfdefault}{bx}{n}

\DeclareMathOperator*{\argmin}{arg\,min}

\DeclareMathOperator*{\minimize}{\text{minimize}}

\newcommand{\btheta}{{\boldsymbol{\theta}}}

\newcommand{\bx}{\mathbf{x}}

\newcommand{\bphi}{\boldsymbol\phi}
\newcommand{\bpsi}{\boldsymbol\psi}

\newcommand{\bz}{\mathbf{z}}

\newcommand{\Def}[0]{\mathrel{\mathop:}=}

\usepackage{pifont}

\usepackage[dvipsnames]{xcolor}

\usepackage{color, colortbl}
\definecolor{Gray}{gray}{0.93}
\definecolor{Orange}{rgb}{1,0.5,0}
\definecolor{DGray}{gray}{0.83}
\definecolor{LightCyan}{rgb}{0.88,1,1}
\definecolor{styleblue}{HTML}{504099}
\definecolor{mypurple}{HTML}{9391ff}
\definecolor{darkgreen}{rgb}{0.00784, 0.5, 0.12549}

\definecolor{WarnREd}{rgb}{1,0.4,0.4}
\definecolor{WarnOrange}{rgb}{1,0.682,0.502}
\definecolor{WarnPink}{rgb}{0.9176, 0.7215, 0.7215}
\definecolor{GoodGreen}{rgb}{0.5019, 0.9215, 0.6039}

\usepackage{soul}

\usepackage[T1]{fontenc}

\definecolor{lightblue}{RGB}{200,220,255}    
\definecolor{lightgreen}{RGB}{220,255,220}   
\definecolor{lightyellow}{RGB}{255,255,200}  
\definecolor{lightred}{RGB}{255,200,200}     
\newcommand{\ours}{\textsc{Water4MU}}

\newcommand{\hidden}{\textsc{HiDDeN}}

\everydisplay{\small}

\def\paperID{15075} 
\def\confName{ICCV}
\def\confYear{2025}

\title{Invisible Watermarks, Visible Gains: \\Steering Machine Unlearning with Bi-Level Watermarking Design}

\author{Yuhao Sun$^{1,2}$\thanks{This work was conducted during a remote research visit to the OPTML Group at Michigan State University.}, Yihua Zhang$^{3}$, Gaowen Liu$^{4}$, Hongtao Xie$^{1}$, Sijia Liu$^{3}$\\
$^1$University of Science and Technology of China, \\~~$^2$Institute of Artificial Intelligence, Hefei Comprehensive National Science Center, \\~~$^3$Michigan State University, ~~$^4$Cisco Research\\
}

\begin{document}
\maketitle

\begin{abstract}
With the increasing demand for the right to be forgotten, machine unlearning (MU) has emerged as a vital tool for enhancing trust and regulatory compliance by enabling the removal of sensitive data influences from machine learning (ML) models. However, most MU algorithms primarily rely on in-training methods to adjust model weights, with limited exploration of the benefits that data-level adjustments could bring to the unlearning process. To address this gap, we propose a novel approach that leverages digital watermarking to facilitate MU. By integrating watermarking, we establish a controlled unlearning mechanism that enables precise removal of specified data while maintaining model utility for unrelated tasks. We first examine the impact of watermarked data on MU, finding that MU effectively generalizes to watermarked data. Building on this, we introduce an unlearning-friendly watermarking framework, termed {\ours}, to enhance unlearning effectiveness. The core of {\ours} is a bi-level optimization (BLO) framework: at the upper level, the watermarking network is optimized to minimize unlearning difficulty, while at the lower level, the model itself is trained independently of watermarking. 
Experimental results demonstrate that {\ours} is effective in MU across both image classification and image generation tasks. Notably, it outperforms existing methods in challenging MU scenarios, known as ``challenging forgets''.
%
\end{abstract}    
\section{Introduction}
\label{sec:intro}
Machine unlearning (MU), which aims to remove the influence of unwanted data from a pre-trained model while preserving the model’s utility on data unrelated to the unlearning request, has emerged as a promising approach for customizing and adapting machine learning (ML) models to diverse contexts and requirements \cite{bourtoule2021machine,nguyen2022survey,triantafillou2024we,liu2024rethinking}.
For example, MU is widely used to enhance model privacy, aligning with the right to be forgotten by removing private or copyrighted information from models to prevent privacy breaches \cite{conley2010right,cao2015towards,tang2023ensuring,dou2024avoiding,zhang2024unlearncanvas}. It has also been applied to improve model generalization in transfer learning by removing the influence of undesired source data in pre-trained models, where removing non-salient source data to downstream tasks improves fine-tuning performance \cite{jain2023data,jia2023model}. Furthermore, MU shows promise in enhancing model robustness, for instance, by removing the influence of backdoored training data to defend against backdoor attacks \cite{liu2022backdoor,wu2024unlearning} or eliminating harmful data to improve the model's trust \cite{liu2024towards,gandikota2023erasing,zhang2024defensive}.
\begin{figure}[t]
    \centering
    \includegraphics[width=0.8\linewidth]{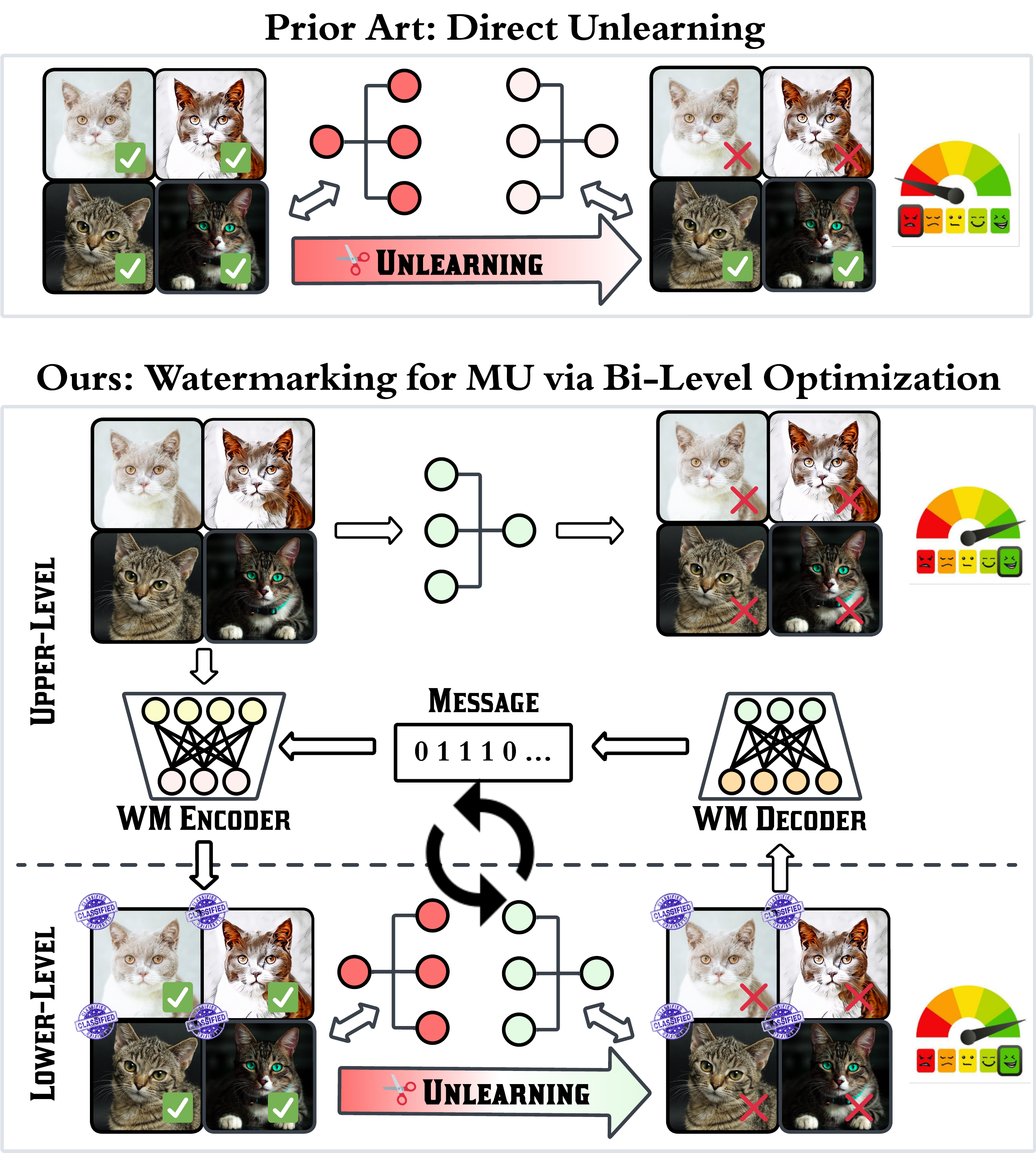}
    \vspace*{-1em}
    \caption{An overview of the watermarking for machine unlearning framework via bi-level optimization proposed in this work.}
    \label{fig: teaser}
    \vspace*{-2em}
\end{figure}

The expanding applications of MU across diverse domains and model types--including discriminative and generative models  \cite{xu2024machine,liu2024machine}--are driven by its effectiveness in erasing specific data influence within ML models. That is, a key aspect of MU involves characterizing the influence of specific data on the model and leveraging this data-model interaction to adjust the model for effective unlearning. However, predominant research in MU focuses on model-based weight updating as the primary approach to achieve the unlearning objective \cite{izzo2021approximate,golatkar2020eternal,becker2022evaluating,thudi2022unrolling,warnecke2021machine,kurmanji2024towards,fan2023salun}, with less attention given to the impact of data modifications--such as \textit{watermarking}, which we will investigate in this work--on MU effectiveness.

Digital watermarking, which embeds ownership signatures (known as watermark messages) into digital content such as images, serves as a valuable tool for tracking data origin and deterring unauthorized use or distribution \cite{xuehua2010digital,zhang2024editguard,hosny2024digital}. Despite the variety of watermarking techniques \cite{shih2003combinational,bhowmik2016visual,zhu2018hidden,luo2020distortion,singh2024digital}, they provide an effective means to \textit{proactively} modify data points in ways that are imperceptible to human users. Deep learning-based watermarking techniques \cite{zhu2018hidden,zhong2023brief}, in particular, offer automated methods to embed and extract watermark messages from watermarked data.
If we consider watermarking as a (privacy-preserving) data modification operation, intriguing questions arise on how it influences unlearning process and whether it can be strategically controlled to facilitate unlearning. Thus,  we ask:
\begin{center}
    \textit{(Q) What is the impact of watermarking on unlearning, and how can it be guided to facilitate unlearning?}
\end{center}

To address \textit{(Q)}, we focus on watermarking for imagery data and MU in the context of image classification. We consider two paradigms for integrating MU with watermarking.
First, in the paradigm of MU for watermarking, we investigate the generalization of the exact unlearning approach, where the image classifier is retrained from scratch with the data samples to be forgotten excluded. We examine how this approach generalizes to watermarked training samples (used in the unlearning optimization process) and watermarked testing samples (used in unlearning evaluation).
We find that unlearning generalizes well to watermarked data; watermarking appears to be orthogonal to unlearning and has minimal negative impact on its effectiveness.
Inspired by this, we consider the second paradigm of watermarking for MU, where we investigate whether the watermarking design can be customized to favor MU. This allows us to proactively watermark data in a way that facilitates the unlearning optimization process. The proposed MU-aware watermarking method is referred to as {\ours}. We provide a schematic overview of {\ours} in Fig.\,\ref{fig: teaser}. In what follows, we summarize \textbf{our contributions}. 

$\bullet$  To the best of our knowledge, this is the first work to investigate the influence of watermarking on MU and to explore their interactions.

$\bullet$ We develop {\ours} using bi-level optimization, enabling the watermarking mechanism to enhance MU by designing the watermarking network and selecting watermark messages to support unlearning when requested.

$\bullet$ We validate the advantages of incorporating watermarking into MU for image classification and generation by comparing the performance of {\ours} with existing MU methods in their standard forms.

\section{Related Work}
\label{sec:related work}
\noindent\textbf{Machine unlearning (MU).}
MU focuses on modifying ML models to remove the influence of specific data points, classes, or broader knowledge-level concepts \cite{cao2015towards,ginart2019making,neel2021descent,sekhari2021remember,ullah2021machine}. A widely-recognized exact unlearning approach, termed Retrain, involves retraining the model from scratch after omitting the data points to be forgotten \cite{thudi2022unrolling,thudi2022necessity}. While Retrain provides the ideal removal of data influences \cite{dwork2006our,graves2021amnesiac}, its computational cost and lack of scalability render it impractical for large-scale deep learning models. To address these limitations, numerous studies have proposed approximate unlearning methods that aim to achieve efficient unlearning without requiring full model retraining \cite{fan2023salun,ginart2019making, golatkar2020eternal,guo2019certified,izzo2021approximate,jia2023model,neel2021descent,sekhari2021remember,thudi2022unrolling,ullah2021machine,warnecke2021machine}.

MU for image classification has predominantly centered on two scenarios: (random) data-wise forgetting and class-wise forgetting. The former involves eliminating the influence of randomly chosen data points from the training set, while the latter aims to remove the effects of an entire image class.
As MU research has evolved, its applications have also extended beyond image classification to encompass other domains. For example, MU has been applied to erasing specific concepts from diffusion models \cite{fan2023salun,gandikota2023erasing,gandikota2024unified,heng2023selective,kumari2023ablating,zhang2023forget,zhang2024unlearncanvas}. Beyond vision tasks, MU is increasingly recognized for its potential to enhance the trustworthiness of data-model relationships in other fields, such as federated learning \cite{che2023fast,wang2022federated,wu2022federated} and large language models (LLMs) \cite{eldan2023whos,wu2023depn,yao2023large,yu2023unlearning}. 

\noindent\textbf{Digital watermarking.}
Digital watermarking serves a wide range of purposes, including copyright protection, content authentication, and data provenance tracking \cite{cox2007digital,miller2000informed,voloshynovskiy2001attacks}. It embeds hidden information, known as a watermark message, into digital images in a way that is imperceptible to human observers but can be extracted using specific algorithms. Traditional watermarking techniques have primarily focused on transform-domain methods, such as discrete cosine transform (DCT) \cite{piva1997dct,barni1998dct}, discrete wavelet transform (DWT) \cite{xia1997multiresolution}, and singular value decomposition (SVD) \cite{liu2002svd}, embedding watermarks within the frequency or singular value domains of images.
Recent advances in ML have driven the development of deep learning-based watermarking techniques. One prominent example is {\hidden} \cite{zhu2018hidden}, which uses an encoder-decoder framework to efficiently embed and extract watermarks from images. Expanding on this approach, methods like DeepStego \cite{zhang2019steganogan} leverage adversarial learning to improve the visual fidelity of watermarked images while maintaining robust watermark extraction. Although prior studies have examined how watermarked data affects various visual tasks \cite{wu2024watermarks,yao2024hide}, the relationship between digital watermarking and MU has remained largely unexplored. This work seeks to address this gap by investigating how watermarking can be utilized to support and enhance MU processes, offering new insights into their interaction and potential synergies.

\noindent\textbf{Visual prompting.} Visual prompting (VP) was initially introduced by \citep{bahng2022visual, jia2022visual} as an adaptation of prompting methods from natural language processing (NLP) to computer vision. A closely related concept, adversarial reprogramming or model reprogramming, was previously applied in computer vision to repurpose fixed pretrained models for new tasks \citep{elsayed2018adversarial, chen2022model, neekhara2018adversarial, neekhara2022cross, chen2021adversarial, zhang2022fairness, chen2022visual,YangGW0CZFFDWX24}. Recent advancements have refined VP techniques by optimizing label mappings \citep{chen2021adversarial, yang2023visual} and introducing improved normalization methods \citep{wu2022unleashing}. The versatility of VP has also been demonstrated across various applications, including adversarial defense \citep{chen2023visual, mao2022understanding}, enhancing differential privacy \citep{li2023exploring}, advancing model sparsification \citep{jin2023visual}, improving model generalization \citep{ma2024visual, zhang2024visual}, and addressing distributional shifts \citep{huang2023diversity, tsai2023self}. Furthermore, VP has been effectively extended to vision-language models, enabling better integration and adaptation for multimodal tasks \citep{zhou2022learning, wu2024visual, kunananthaseelan2024lavip}.
\section{Preliminaries and Problem Statement}
\label{sec:preliminaries}


\noindent\textbf{MU setup and formulation.} 
The \textit{goal} of MU is to remove the influence of specific (undesired) training data from a pre-trained ML model while retaining its normal model utility. To formalize, let $\mathcal{D} = 
\{\bz_i\}_{i=1}^N$ denote the training dataset with $N$ data points, where each data point includes a feature vector $\bx_i$ and label $y_i$ in a supervised learning setup. We define $\Df \subseteq \mathcal{D}$ as the dataset subset we aim to forget, termed the \textit{forget set}, and let $\Dr = \mathcal{D} \setminus \mathcal{D}_f$ denote the \textit{retain set}, comprising the data to be preserved.

Retraining the model from scratch solely on the retain set $\mathcal{D}_r$ is generally regarded as the gold standard in MU, as it minimizes privacy leakage of the forget data in $\mathcal{D}_f$ post-retraining and is considered an \textit{exact} unlearning approach \cite{thudi2022unrolling,jia2023model,georgiev2024attribute}. 
However, retraining is computationally intensive and often impractical. Consequently, current research has focused on \textit{approximate} unlearning methods \citep{golatkar2020eternal, becker2022evaluating, chen2023boundary, warnecke2021machine,triantafillou2024we}, which aim to achieve efficient unlearning directly from an already-trained model, without the need for retraining from scratch.
Let $\btheta_\mathrm{o}$ denote the original model trained on the full dataset $\mathcal{D}$ using, for instance, empirical risk minimization.
Approximate unlearning can be framed as a fine-tuning problem, as outlined below:
{\small
\begin{align}
\displaystyle \btheta_{\mathrm{u}} \Def \argmin_\btheta ~ 
&\mathcal{L}_{\mathrm{mu}} (\btheta; \Df, \Dr) \nonumber \\
&\Def 
\lambda_{\mathrm{f}}\ell_{\mathrm{f}}(\btheta; \Df)+\lambda_{\mathrm{r}}  \ell_{\mathrm{r}}(\btheta;\Dr),
\label{eq: MU_prob}
\end{align}
}%
where  $\btheta_{\mathrm{u}}$ represents the unlearned model (\textit{i.e.}, the updated model after unlearning from the original model $\btheta_{\mathrm{o}}$). 

In \eqref{eq: MU_prob},
the terms $\ell_{\mathrm{f}}$ and $\ell_{\mathrm{r}}$ represent the forget and retain losses, respectively. The former characterizes the unlearning objective on the forget set, often defined as the \textit{negative} of the cross-entropy (CE) based prediction loss ($\ell_{\mathrm{CE}}$)  to effectively reverse the influence of the forget data. The latter, analogous to standard ERM, enforces the preservation of the model's utility on the retain set, ensuring that performance on unaffected data remains intact.
There is generally a tradeoff between unlearning effectiveness (\textit{i.e.}, minimizing $\ell_{\mathrm{f}}$) and preserving model utility (\textit{i.e.}, minimizing  $\ell_{\mathrm{r}}$). To balance this,    regularization parameters $\lambda_{\mathrm{f}} \geq 0$ and  $\lambda_{\mathrm{r}} \geq 0$ are introduced in \eqref{eq: MU_prob}  for controlled adjustment between these objectives.
%
If we set $\lambda_{\mathrm{f}} = 1$, $\lambda_{\mathrm{r}} = 0$, and $\ell_{\mathrm{f}} = -\ell_{\mathrm{CE}}$, then solving the problem reduces to the classic gradient ascent ({GA}) approach for MU  \cite{thudi2022unrolling}.
If we set $\lambda_{\mathrm{f}} = 0$, $\lambda_{\mathrm{r}} = 1$, and $\ell_{\mathrm{r}} = \ell_{\mathrm{CE}}$, then problem \eqref{eq: MU_prob} reduces to the classic fine-tuning ({FT}) approach, which leverages catastrophic forgetting to achieve unlearning  \cite{golatkar2020eternal}.

\noindent\textbf{Image watermarking.} Digital watermarking is another fundamental tool employed in this paper, which
%
involves embedding hidden information (referred to as the ``watermark message'') into original imagery data through a watermarking process. 
This then allows for the watermark message to be extracted through a decoding process. 
Such an encoding-decoding watermarking technique has been used to assert data ownership and authenticate content \cite{xuehua2010digital,zhang2024editguard,hosny2024digital}. 

Formally, the watermarking problem can be defined as follows: Given an input image $\mathbf{x}$ and an $L$-bit watermark message $\mathbf{m} \in \{0,1\}^L$, the objective is to generate a watermarked image $\mathbf{x}_{\mathrm{w}}$ that preserves visual similarity to the original image $\mathbf{x}$ while embedding the watermark message $\mathbf{m}$. Additionally, the watermarked image $\mathbf{x}_{\mathrm{w}}$ should be constructed to allow reverse engineering of its embedded message $\mathbf{m}$, thereby enabling the identification of the image’s origin.
To this end, we adopt {\hidden} \cite{zhu2018hidden} as the watermarking method--a widely used deep neural network-based image watermarking framework. It comprises an encoder $f_{\boldsymbol{\psi}}$ and a decoder $g_{\bphi}$, with network parameters $\boldsymbol{\psi}$ and $\bphi$, respectively. The encoder $f_{\boldsymbol{\psi}}$ takes the original image $\bx$ and the message $\mathbf{m}$ as input, producing the watermarked image $\mathbf{x}_{\mathrm{w}} = f_{\boldsymbol{\psi}}(\bx, \mathbf{m})$. The decoder $g_{\bphi}$ then aims to recover the watermark message from $\mathbf{x}_{\mathrm{w}}$, yielding an extracted message denoted as $\hat{\mathbf m} = g_{\bphi}(\mathbf{x}_{\mathrm{w}})$.
The encoder and decoder of {\hidden} are then trained jointly to minimize both the image reconstruction loss ($\ell_{\mathrm{rec}}$) between $\mathbf{x}_{\mathrm{w}}$ and $\mathbf{x}$, ensuring visual similarity, and the watermark message decoding loss ($\ell_{\mathrm{dec}}$), which ensures accurate recovery of the embedded message.
This yields the watermark network training problem below  
{\small
\begin{align}
    {\boldsymbol{\psi}}_{\mathrm{w}},\bphi_{\mathrm{w}} 
    = \argmin_{{\boldsymbol{\psi}},\bphi}~ & \mathcal{L}_{\mathrm{wm}} (\boldsymbol{\psi}, \boldsymbol{\phi}; \mathbf m, \mathcal{D})  \nonumber \\
  &  \hspace*{-11mm} \Def   \mathbb E_{\mathbf x \in \mathcal D} [ \ell_{\mathrm{rec}} (\mathbf{x}_{\mathrm{w}} ,\bx)  + \ell_{\mathrm{dec}}(g_{\bphi}(\mathbf{x}_{\mathrm{w}}), \mathbf m)  ],
  \hspace*{-3mm}
\label{eq: watermark_problem}
\end{align}
}%
where ${\boldsymbol{\psi}}$ and $\bphi$ are encoding and decoding network parameters, 
$\mathcal{D}$ denotes the dataset,  and $\bx_{\mathrm{w}}=f_{\boldsymbol{\psi}}(\bx,\mathbf m)$.
Once the encoding network is trained by solving \eqref{eq: watermark_problem}, we can watermark data by applying $  f_{\boldsymbol{\psi}_{\mathrm{w}}}(\bx, \mathbf{m})$ to a given data point $\mathbf{x}$ with the specified watermark message $\mathbf{m}$.

\noindent\textbf{Problem of interest: MU meets watermarking.}
Our study investigates the synergy between MU and watermarking, serving two primary purposes: \textit{(P1) MU for watermarking}, which explores the generalizability of MU in unlearning watermarked data;
\textit{(P2) Watermarking for MU}, which examines how watermarking can be strategically designed to facilitate effective unlearning.

The rationale behind investigating \textit{(P1)} is that watermarking can be viewed as a form of data shift, as it introduces alterations to the original dataset. This prompts an examination of whether MU remains generalizable to such shifts during both evaluation and training. 
%
Building on the insights gained from exploring \textit{(P1)}, we pose \textit{(P2)} to examine whether adjustments to the watermarking network in \eqref{eq: watermark_problem} and the selection of the watermark message $\mathbf{m}$ can proactively influence unlearning performance. Here, we aim for a synergistic relationship where watermarking facilitates the MU process, creating a mutually beneficial framework for data watermarking and MU.

\section{Generalization of MU to Watermarked Data}
\label{sec:general}
In this section, we address \textit{(P1)} by examining how an off-the-shelf watermarking technique like {\hidden}, when applied to both forget and retain sets, impacts unlearning during training and evaluation.
Specifically, given an unlearned model $\btheta_{\mathrm{u}}$ obtained by solving the unlearning problem \eqref{eq: MU_prob} on the original (unwatermarked) forget set $\Df$, we first assess if the unlearning effectiveness of $\btheta_{\mathrm{u}}$ generalizes to the watermarked forget set $\hat{\Df} \Def \{ \mathbf x_{\mathrm{w}}  \}_{\mathbf x \in \Df}$ and the watermarked retain or testing set at the evaluation phase, where recall that $\mathbf{x}_{\mathrm{w}} = f_{\boldsymbol{\psi}_{\mathrm{w}}}(\bx, \mathbf{m})$ as defined in \eqref{eq: watermark_problem}.
Similarly, if the MU problem \eqref{eq: MU_prob} is performed on the watermarked dataset, we aim to investigate how its unlearning performance generalizes when evaluated on the \textit{unwatermarked} forget set.
The investigation above can be summarized into the following two scenarios of interest ($\mathcal{S}_1$-$\mathcal S_2$):%

\noindent$\bullet$ \textbf{($\mathcal{S}_1$) Unwatermarked unlearning with watermarked evlauation}: Standard unlearning optimization is performed on the \textit{unwatermarked} dataset, as in \eqref{eq: MU_prob}, while evaluation is conducted on the \textit{watermarked} MU evaluation sets.

\noindent$\bullet$ \textbf{($\mathcal{S}_2$) Watermarked unlearning with unwatermarked evaluation}: Unlearning optimization is applied to the \textit{watermarked} forget and retain sets using the {\hidden} watermarking network ($f_{\boldsymbol{\psi}_{\mathrm{w}}}(\bx, \mathbf{m})$), while evaluation is performed on the \textit{unwatermarked} datasets.

Both cases will be validated against the baseline scenario: \textbf{($\mathcal{S}_0$) unwatermarked unlearning with unwatermarked evaluation}.
Our rationale is that comparing $\mathcal{S}_1$ to $\mathcal{S}_0$ reveals MU’s robustness to data shifts introduced by watermarking at the testing stage, while comparing $\mathcal{S}_2$ to $\mathcal{S}_0$ highlights the impact of watermarking on unlearning training.

To examine the proposed scenarios, we focus on the exact unlearning method of retraining from scratch on the retain set only, referred to as ``Retrain''. We did not prioritize approximate unlearning methods to avoid the confounding influence of approximation factors when analyzing the watermarking effect in MU. While we acknowledge that Retrain may not be scalable for large datasets, it provides a controlled setting to draw precise insights into the impact of watermarking. To measure \textit{unlearning effectiveness}, we follow \cite{jia2023model} and use two metrics: unlearning accuracy ({UA}), {defined as $1 -$ (accuracy of the unlearned model $\btheta_{\mathrm{u}}$ on the forget set)}, and MIA-Efficacy, which measures the prediction accuracy of a membership inference attack (MIA)-based detector in correctly identifying the forget data as non-training samples.
To assess \textit{model utility} after unlearning, we use retain accuracy (RA) and testing accuracy (TA), which evaluate the prediction accuracy of $\btheta_{\mathrm{u}}$ on  $\Dr$ and the testing set, respectively.

\begin{figure}[t]
    \centering
\includegraphics[width=0.7\linewidth]{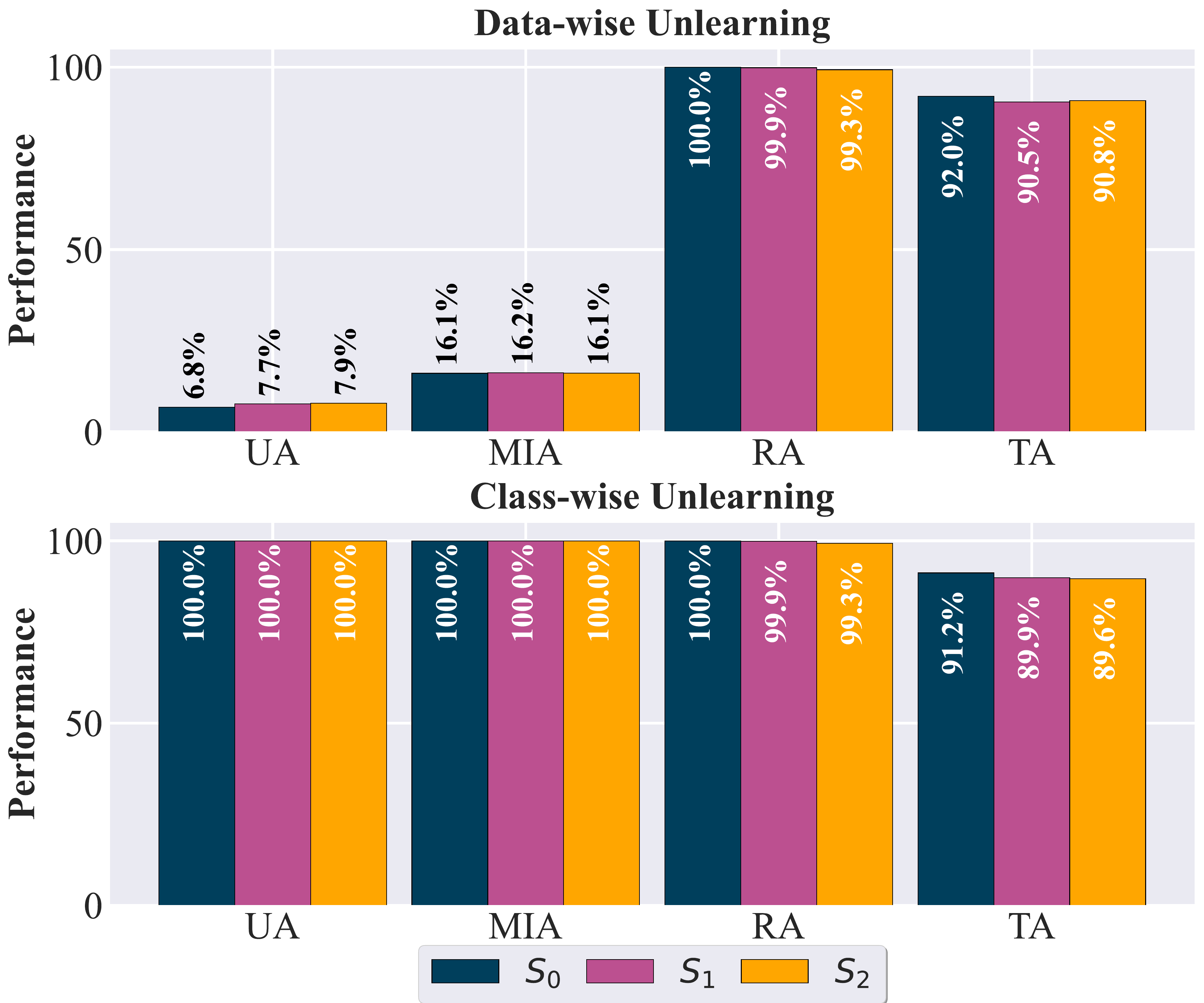}
    \vspace*{-0.8em}
    \caption{Performance evaluation of the exact unlearning method, Retrain, across the studied MU-watermarking interaction scenarios: $\mathcal{S}_1$ (unwatermarked unlearning with watermarked evaluation) and $\mathcal{S}_2$ (watermarked unlearning with unwatermarked evaluation), compared to the unwatermarked baseline $\mathcal{S}_0$.
    }
    \vspace*{-1.5em}
    \label{fig: data_class_comparison}
\end{figure}

Given the evaluation metrics (UA, MIA-Efficacy, RA, TA), \textbf{Fig.\,\ref{fig: data_class_comparison}}
presents the performance of Retrain under different MU-watermarking interaction scenarios ($\mathcal{S}_1$ and $\mathcal{S}_2$) vs. the unwatermarked baseline $\mathcal{S}_0$ in both data-wise and forget-wise forgetting. 
As illustrated by the comparison of $\mathcal{S}_1$ vs. $\mathcal{S}_0$, the application of watermarking to unlearning evaluation sets shows negligible impact on the effectiveness of MU, demonstrating its robustness against data modifications introduced by watermarking. Furthermore, in the comparison of $\mathcal{S}_2$ vs. $\mathcal{S}_0$, performing unlearning optimization directly on watermarked datasets also demonstrates minimal impact on the unlearning outcomes. This finding highlights the seamless compatibility of MU methods with watermarked data, ensuring reliable and consistent performance across both standard and watermarked scenarios.
This observation further motivates our investigation in the next section, where we explore proactive watermarking design to enhance and facilitate the MU process.





\section{{\ours}: MU Enhanced by Watermark}
\label{sec:method}

In this section, we address \textit{(P2)}, as described in Sec.\,\ref{sec:preliminaries}, to  investigate how the watermarking mechanism in \eqref{eq: watermark_problem} can be strategically designed to proactively facilitate MU when the watermark is applied to the forget data.
We refer to the proposed approach as {\ours}.

\noindent\textbf{A bi-level optimization (BLO) view of {\ours}.}
BLO is used to characterize a leader-follower game \cite{zhang2024introduction}, in which the leader’s actions depend on the optimal solution determined by the follower. 
If we consider the watermarking procedure conducted by the leader and the unlearning procedure by the follower, we can apply BLO to design the watermarking process in a way that facilitates the unlearning process.
Specifically, applying BLO involves two levels of optimization: the \textit{upper-level} optimization, corresponding to the leader's task (\textit{i.e.}, optimization for watermarking), and the \textit{lower-level} optimization, corresponding to the follower's task (\textit{i.e.}, optimization for unlearning).
In the following, we detail each optimization level in our proposal, {\ours}.

We start with the design of the lower-level optimization for unlearning, given the watermarked datasets. Considering the unlearning objective $\mathcal{L}_{\mathrm{mu}}$ as formulated in \eqref{eq: MU_prob}, with watermarked data produced by the watermarking encoder $f_{\boldsymbol{\psi}}$ in \eqref{eq: watermark_problem}, the follower in the lower-level optimization seeks to obtain the unlearned model by minimizing   $\mathcal{L}_{\mathrm{mu}}$ over the watermarked forget and retain sets:
{\small
\begin{align}
 \btheta_\mathrm{u}(\bpsi)=\argmin_\btheta ~
 \underbrace{-\ell_\mathrm{CE}(\btheta;\hat{\mathcal D_\mathrm{f}})+ \mathcal \ell_\mathrm{CE}(\btheta;\hat{\mathcal D_\mathrm{r}})}_\text{$= \mathcal{L}_{\mathrm{mu}} (\btheta; \hat{\mathcal D_\mathrm{f}}, \hat{\mathcal D_\mathrm{r}})$},  
\label{eq: lower_level_MU}
\end{align}
}%
where $\hat{\mathcal D_\mathrm{f}}$ (or $\hat{\mathcal D_\mathrm{r}}$) denotes the watermarked forget (or retain) dataset, \textit{i.e.}, $\hat{\Df} \Def \{ \mathbf x_{\mathrm{w}}  \}_{\mathbf x \in \Df}$ and $\hat{\Dr} \Def \{ \mathbf x_{\mathrm{w}}  \}_{\mathbf x \in \Dr}$ with $\bx_{\mathrm{w}}=f_{\boldsymbol{\psi}}(\bx,\mathbf m)$ in \eqref{eq: watermark_problem}.
In \eqref{eq: lower_level_MU}, we explicitly express the minimizer $\btheta_{\mathrm{u}}(\bpsi)$ as a function of the watermarking network parameters $\bpsi$.
In addition,  unless specified otherwise, we define the unlearning objective of \eqref{eq: lower_level_MU} by setting $\lambda_{\mathrm{f}} = \lambda_{\mathrm{r}} = 1$, $\ell_{\mathrm{r}} = \ell_{\mathrm{CE}}$, and $\ell_{\mathrm{f}} = -\ell_{\mathrm{CE}}$ in \eqref{eq: MU_prob}. This setup is commonly known as gradient difference method (GradDiff)~\cite{maini2024tofu,liu2024rethinking}.


Next, we introduce the upper-level optimization for the MU-aware watermarking design. Given the lower-level problem \eqref{eq: lower_level_MU}, the upper-level optimization has two primary objectives: \textit{(a)} ensuring that the lower-level unlearned model $\btheta_{\mathrm{u}}(\bpsi)$, trained on watermarked datasets, performs effectively when evaluated on the original unwatermarked datasets, and \textit{(b)} maintaining the effectiveness of the watermarking network to reliably encode and decode the watermark message.
This leads to the following  objective:
{\small
\begin{align}
    \hat{\mathcal{L}} (\boldsymbol{\psi}, \boldsymbol{\phi},  \btheta_\mathrm{u}(\bpsi)) \Def & \underbrace{\mathcal{L}_{\mathrm{mu}} (\btheta_\mathrm{u}(\bpsi); \Df, \Dr)}_\text{(a) Unlearning validation} \nonumber \\
    & + \underbrace{
    \mathcal{L}_{\mathrm{wm}} (\boldsymbol{\psi}, \boldsymbol{\phi}; \mathbf m, \Df \cup \Dr)}_\text{(b) Watermarking validation},
    \label{eq: upper_level_obj}
\end{align}
}%
where $\mathcal{L}_{\mathrm{mu}}$ is defined in \eqref{eq: MU_prob} with the GradDiff specification as in \eqref{eq: lower_level_MU} in order to validate the lower-level unlearned model $\btheta_{\mathrm{u}}(\bpsi)$  on the original MU dataset,  and $\mathcal{L}_{\mathrm{wm}}$ represents the training loss of the watermarking network, as defined in \eqref{eq: watermark_problem}.

Integrating \eqref{eq: upper_level_obj} with \eqref{eq: lower_level_MU} forms the proposed BLO problem required for the design of {\ours}. It is evident from \eqref{eq: upper_level_obj} that these two optimization levels are interconnected through the upper-level watermarking encoder parameters $\boldsymbol{\psi}$, which influence the lower-level solution $\btheta_{\mathrm{u}}(\bpsi)$. Therefore,   the BLO problem for {\ours} is cast as 
{\small
\begin{align}
    \begin{array}{ll}
\displaystyle \minimize_{\boldsymbol{\psi}, \boldsymbol{\phi}}         &      \hat{\mathcal{L}} (\boldsymbol{\psi}, \boldsymbol{\phi},  \btheta_\mathrm{u}(\bpsi))  \\
     \text{subject to}    &  \btheta_\mathrm{u}(\bpsi)=\argmin_\btheta ~\mathcal{L}_{\mathrm{mu}} (\btheta; \hat{\mathcal D_\mathrm{f}}, \hat{\mathcal D_\mathrm{r}}),
    \end{array}
    \label{eq: BLO_Water4MU}
\end{align}}%
where $\hat{\mathcal{L}} $ and $\mathcal{L}_{\mathrm{mu}}$ have been defined in \eqref{eq: lower_level_MU} and \eqref{eq: upper_level_obj} respectively, and recall that the watermarked datasets $\hat{\mathcal D_\mathrm{f}}$ and $\hat{\mathcal D_\mathrm{r}}$ depend on the upper-level watermarking encoder  $\boldsymbol{\psi}$.

\noindent\textbf{Solving problem \eqref{eq: BLO_Water4MU} via implicit gradient.}
To solve the BLO problem \eqref{eq: BLO_Water4MU}  using a standard gradient descent approach to minimize the upper-level objective function, we face a difficulty known as the implicit gradient (IG) challenge. This arises from the gradient of the upper-level objective, which depends implicitly on the lower-level solution. Specifically, the gradient of the upper-level objective involves: 
{\small\begin{align}
\raisetag{6mm}
\text{\resizebox{0.9\linewidth}{!}{$
\begin{array}{c}
     \hspace*{-3mm}
    \frac{d     \hat{\mathcal{L}} }{ d \boldsymbol{\psi}}  = \nabla_{\boldsymbol{\psi}}\hat{\mathcal{L}} (\boldsymbol{\psi}, \boldsymbol{\phi},  \btheta_\mathrm{u} ) + \underbrace{\frac{d \btheta_\mathrm{u}(\bpsi) }{d \boldsymbol{\psi}} }_\text{IG} \nabla_{\btheta}\hat{\mathcal{L}} (\boldsymbol{\psi}, \boldsymbol{\phi},  \btheta ) \left. \right|_{\btheta = \btheta_\mathrm{u}},
\end{array}
$}}
\label{eq: upper_grad}
\end{align}}%
where $\nabla_{\boldsymbol{\psi}}\hat{\mathcal{L}}   $  or  $\nabla_{\boldsymbol{\theta}}\hat{\mathcal{L}}   $ denotes the partial derivative ($\frac{\partial \cdot}{\partial \cdot}$) of the multi-variate function
$\hat{\mathcal{L}} (\boldsymbol{\psi}, \boldsymbol{\phi},  \btheta  )$, and $\frac{d \cdot}{ d \cdot}$ denotes the full derivative. 
In \eqref{eq: upper_grad}, $\frac{d \btheta_\mathrm{u}(\bpsi)}{d \boldsymbol{\psi}}$ represents the  IG, as $\btheta_\mathrm{u}(\bpsi)$ is obtained by solving a coupled but implicit lower-level optimization problem in \eqref{eq: BLO_Water4MU}, making its closed-form solution challenging to derive.

To address the IG challenge, we employ a widely used optimization technique known as the \textit{implicit function} approach \cite[Sec.\,III-A]{zhang2024introduction}. This approach leverages the Implicit Function Theorem \cite{krantz2002implicit}, assuming that a singleton solution exists for the lower-level problem.
This leads to the following expression of IG: 
{\small
\begin{align}
\frac{d \btheta_\mathrm{u}(\bpsi) }{d \boldsymbol{\psi}} =-\nabla^2_{\bpsi\btheta} \ell_{\mathrm{mu}} \, (\nabla^2_{\btheta\btheta}\ell_{\mathrm{mu}})^{-1},
\label{eq: IG_IF}
\end{align}}%
where $\ell_{\mathrm{mu}}$ is the lower-level objective function of 
\eqref{eq: BLO_Water4MU}, we have used and will continue to use omitted input arguments in functions for brevity, 
$\nabla^2_{\bpsi\btheta}$
denotes the (cross-variable) second-order partial derivative, $\nabla^2_{\btheta\btheta}$ is the Hessian matrix, and $^{-1}$ is the matrix inversion. 
To further simplify the IG expression in \eqref{eq: IG_IF}, a common approximation assumes a diagonal matrix for the Hessian. This assumption is reasonable because the prediction loss landscape of deep neural networks is often considered piecewise linear in a tropical hyper-surface \cite{alfarra2022decision}.
 By setting $\nabla^2_{\btheta\btheta} \ell_{\mathrm{mu}} = \lambda \mathbf{I}$, where $\lambda$ is a small hyperparameter to adjust the approximation, we can simplify \eqref{eq: IG_IF} to:
{\small
\begin{align}
\frac{d \btheta_\mathrm{u}(\bpsi) }{d \boldsymbol{\psi}} =- (1/\lambda) \nabla^2_{\bpsi\btheta} \ell_{\mathrm{mu}} .
\label{eq: IG_IF_Simple}
\end{align}}%
Substituting \eqref{eq: IG_IF_Simple} into \eqref{eq: upper_grad}, the gradient of the upper-level objective becomes
{\small
\begin{align}
    \frac{d     \hat{\mathcal{L}} }{ d \boldsymbol{\psi}}  = \nabla_{\boldsymbol{\psi}}\hat{\mathcal{L}}    -\nabla^2_{\bpsi\btheta} \ell_{\mathrm{mu}}  \nabla_{\btheta}\hat{\mathcal{L}} = \nabla_{\boldsymbol{\psi}}\hat{\mathcal{L}}    - 
    \frac{\partial }{\partial \boldsymbol{\psi}}[\nabla_{\btheta} \ell_{\mathrm{mu}}^\top  \nabla_{\btheta}\hat{\mathcal{L}} ] ,
    \nonumber 
\end{align}}%
where $^\top$ is the transpose of a vector. 
The above implies that the upper-level gradient calculation to perform gradient descent on \eqref{eq: BLO_Water4MU} can be achieved through the first-order derivatives of two scalar-valued functions, $\hat{\mathcal{L}}$ and $\nabla_{\btheta} \ell_{\mathrm{mu}}^T \nabla_{\btheta} \hat{\mathcal{L}}$.
This allows us to implement {\ours} using a standard first-order optimization solver to address \eqref{eq: BLO_Water4MU}.

\noindent\textbf{{\ours} in the image generation context.}
The previously proposed BLO problem was conceived for unlearning in the context of image classification. We can also extend $\ours$ to the prompt-wise forgetting in image generation context. Here, a prompt refers to a text condition used for text-to-image generation, known as a `concept` within MU for generative models \cite{gandikota2023erasing}. In the lower-level optimization, we conduct unlearning on the prompt with corresponding watermarked forget and retain sets. Then, the upper-level optimization ensures that the unlearned model performs well when evaluated on the unwatermarked datasets and maintains the effectiveness of the watermarking network. See Appx.\,\ref{sec:sd} for details.      

\noindent\textbf{Another extended case: Watermark message ($\mathbf m$) selection.}
In the standard watermark network training setup, like in {\hidden}, the watermark message $\mathbf{m}$ is treated as a random message.
However, given a fixed watermarking network, we can also enhance {\ours} by selecting a watermark message $\mathbf{m}$ that facilitates MU. This introduces the problem of watermark message selection while keeping the watermark network parameters $(\boldsymbol{\psi}, \boldsymbol{\phi})$ unchanged.
Analogous to \eqref{eq: BLO_Water4MU}, we substitute the upper-level optimization variables with the watermark message variables $\mathbf{m}$. As shown in \eqref{eq: lower_level_MU}, the watermark message $\mathbf{m}$ influences the lower-level unlearning objective, thereby serving as the new coupling variable that connects the lower-level optimization to the upper-level optimization. 
A similar implicit gradient approach can be applied to solve this extended case. 

\section{Experiments}
\label{sec:experiments}


\subsection{Experiment setups}

\noindent\textbf{Watermarking and unlearning setups.}
For watermarking, we use the pre-trained {\hidden} \cite{zhu2018hidden} as our baseline watermarking network, which we then fine-tune when integrated with unlearning.
For unlearning, we consider three MU scenarios in image classification and image generation.
In the \textit{random data} forgetting scenario, we perform primary experiments on the \textsc{CIFAR-10} dataset using a \textsc{ResNet-18} model, with additional evaluations on \textsc{CIFAR-100} and \textsc{SVHN}. In the \textit{class-wise} forgetting scenario, we focus on the \textsc{CIFAR-10} and \textsc{ImageNet} datasets, also using a \textsc{ResNet-18} model. In prompt-wise forgetting, we use \textsc{UnlearnCanvas}\cite{zhang2024unlearncanvas} dataset with the corresponding fine-tuned Stable Diffusion (SD) v1.5. In addition to standard unlearning tasks, we consider a \textit{worst-case} scenario, known as ``\textit{challenging forgets}'' \cite{fan2025challenging}, which identifies data points that are particularly difficult to unlearn. This approach ranks data based on unlearning difficulty, providing a subset of training samples that present the greatest challenges for unlearning.

\begin{table*}[htb]
\vspace*{-1em}
    \centering 
    \caption{Performance of different unlearning methods under unwatermarked forget/retain sets (Original) and $\ours$-induced watermarked forget/retain sets on (\textsc{CIFAR-10}, \textsc{ResNet-18}). The result format is given by $a_{\pm b}$ with mean $a$ and standard deviation $b$ over 10 independent trials. We present the results under random data forgetting and class-wise forgetting scenarios. The forgetting data of random data forgetting ratio is 10\% of the whole training dataset. The performance difference is provided in Diff.}
   \vspace*{-1em}
    \resizebox{0.95\linewidth}{!}{
    \begin{tabular}{c|ccc|ccc|ccc|ccc|ccc}
        \toprule[1pt]
        \midrule
         \multirow{2}{*}{Metric}&\multicolumn{3}{c|}{\textbf{Retrain}}&\multicolumn{3}{c|}{\textbf{GA}}&\multicolumn{3}{c|}{\textbf{FT}}&\multicolumn{3}{c|}{\textbf{Sparse}}&\multicolumn{3}{c}{\textbf{IU}}\\
         &\multicolumn{1}{c}{Original}&\multicolumn{1}{c}
         {$\ours$}&Diff&\multicolumn{1}{c}{Original}&\multicolumn{1}{c}{$\ours$}&Diff&\multicolumn{1}{c}{Original}&\multicolumn{1}{c}{$\ours$}&Diff&\multicolumn{1}{c}{Original}&\multicolumn{1}{c}{$\ours$}&Diff&\multicolumn{1}{c}{Original}&\multicolumn{1}{c}{$\ours$}&Diff\\
         \midrule
         \multicolumn{16}{c}{10\% random data forgetting}\\
         \midrule
         UA$\uparrow$&$6.78_{\pm0.63}$&$10.01_{\pm0.69}$&\textcolor{darkgreen}{3.23$\blacktriangle$}&$0.80_{\pm0.32}$&$1.92_{\pm0.25}$&\textcolor{darkgreen}{1.12$\blacktriangle$}&$1.85_{\pm0.69}$&$4.93_{\pm0.76}$&\textcolor{darkgreen}{3.08$\blacktriangle$}&$6.11_{\pm0.72}$&$7.50_{\pm0.64}$&\textcolor{darkgreen}{1.39$\blacktriangle$}&$0.64_{\pm0.21}$&$2.62_{\pm0.22}$&\textcolor{darkgreen}{1.98$\blacktriangle$}\\
         MIA$\uparrow$&$16.06_{\pm0.07}$&$19.33_{\pm0.05}$&\textcolor{darkgreen}{3.27$\blacktriangle$}&$1.89_{\pm0.02}$&$5.67_{\pm0.02}$&\textcolor{darkgreen}{3.78$\blacktriangle$}&$5.60_{\pm0.22}$&$8.26_{\pm0.12}$&\textcolor{darkgreen}{2.66$\blacktriangle$}&$13.08_{\pm0.08}$&$14.70_{\pm0.02}$&\textcolor{darkgreen}{1.62$\blacktriangle$}&$1.53_{\pm0.01}$&$3.67_{\pm0.01}$&\textcolor{darkgreen}{2.14$\blacktriangle$}\\
         RA$\uparrow$&$100.00_{\pm0.00}$&$99.93_{\pm0.01}$&\textcolor{blue}{0.07$\blacktriangledown$}&$99.42_{\pm0.27}$&$99.18_{\pm0.34}$&\textcolor{blue}{0.24$\blacktriangledown$}&$99.66_{\pm0.06}$&$98.75_{\pm0.10}$&\textcolor{blue}{0.91$\blacktriangledown$}&$97.76_{\pm0.32}$&$97.22_{\pm0.23}$&\textcolor{blue}{0.54$\blacktriangledown$}&$99.43_{\pm0.22}$&$98.98_{\pm0.36}$&\textcolor{blue}{0.45$\blacktriangledown$}\\

         TA$\uparrow$&$92.02_{\pm0.02}$&$90.63_{\pm0.02}$&\textcolor{blue}{1.39$\blacktriangledown
         $}&$92.19_{\pm0.52}$&$92.39_{\pm0.38}$&\textcolor{darkgreen}{0.20$\blacktriangle$}&$93.54_{\pm0.27}$&$92.11_{\pm0.41}$&\textcolor{blue}{1.43$\blacktriangledown
         $}&$91.61_{\pm0.46}$&$91.65_{\pm0.55}$&\textcolor{darkgreen}{0.04$\blacktriangle$}&$94.51_{\pm0.87}$&$92.87_{\pm0.74}$&\textcolor{blue}{1.64$\blacktriangledown$}\\
         \midrule
         \multicolumn{16}{c}{class-wise forgetting}\\
         \midrule
         UA$\uparrow$&$100.00_{\pm0.00}$&$100.00_{\pm0.00}$&{0.00-}&$41.84_{\pm7.35}$&$49.03_{\pm5.46}$&\textcolor{darkgreen}{7.19$\blacktriangle$}&$37.29_{\pm8.18}$&$53.25_{\pm6.55}$&\textcolor{darkgreen}{15.96$\blacktriangle$}&$87.09_{\pm2.19}$&$94.60_{\pm2.02}$&\textcolor{darkgreen}{7.51$\blacktriangle$}&$92.03_{\pm2.54}$&$98.49_{\pm1.12}$&\textcolor{darkgreen}{6.46$\blacktriangle$}\\
         MIA$\uparrow$&$100.00_{\pm0.00}$&$100.00_{\pm0.00}$&{0.00-}&$55.11_{\pm8.32}$&$61.42_{\pm7.75}$&\textcolor{darkgreen}{6.31$\blacktriangle$}&$55.96_{\pm6.49}$&$69.24_{\pm9.17}$&\textcolor{darkgreen}{13.28$\blacktriangle$}&$90.62_{\pm2.13}$&$98.71_{\pm1.14}$&\textcolor{darkgreen}{8.09$\blacktriangle$}&$99.32_{\pm0.21}$&$100.00_{\pm0.00}$&\textcolor{darkgreen}{0.68$\blacktriangle$}\\
         RA$\uparrow$&$100.00_{\pm0.00}$&$99.84_{\pm0.03}$&\textcolor{blue}{0.16$\blacktriangledown$}&$99.20_{\pm0.12}$&$99.07_{\pm0.23}$&\textcolor{blue}{0.13$\blacktriangledown$}&$99.31_{\pm0.07}$&$98.76_{\pm0.12}$&\textcolor{blue}{0.55$\blacktriangledown$}&$99.58_{\pm0.05}$&$95.27_{\pm0.33}$&\textcolor{blue}{4.31$\blacktriangledown$}&$92.27_{\pm0.26}$&$92.17_{\pm0.41}$&\textcolor{blue}{0.10$\blacktriangledown$}\\

         TA$\uparrow$&$91.20_{\pm0.11}$&$89.95_{\pm0.21}$&\textcolor{blue}{1.25$\blacktriangledown$}&$87.87_{\pm0.07}$&$88.62_{\pm0.18}$&\textcolor{darkgreen}{0.75$\blacktriangle$}&$90.67_{\pm0.22}$&$90.61_{\pm0.14}$&\textcolor{blue}{0.06$\blacktriangledown$}&$90.48_{\pm0.31}$&$90.24_{\pm0.10}$&\textcolor{blue}{0.24$\blacktriangledown$}&$89.50_{\pm0.07}$&$88.67_{\pm0.05}$&\textcolor{blue}{0.83$\blacktriangledown$}\\
         \midrule
         \bottomrule[1pt]
    \end{tabular}}
    \vspace*{-0.8em}
    \label{tab:cifar10_main}
\end{table*}

%

\noindent\textbf{Unlearning baseline methods and evaluation metrics.} For image classification, we compare our proposed {\ours} with the exact unlearning method Retrain and four approximate unlearning methods, FT (fine-tuning) \cite{golatkar2020eternal}, GA (gradient ascent) \cite{thudi2022unrolling}, $\ell_1$ sparsity-promoted unlearning (referred to as Sparse) \cite{jia2023model}, and IU (influence unlearning) \cite{izzo2021approximate}. For image generation, we compare {\ours} with three prompt-wise unlearning methods, erased stable diffusion (ESD)\cite{gandikota2023erasing}, forget-me-not (FMN)\cite{zhang2023forget} and unified concept editing (UCE)\cite{gandikota2024unified}. For evaluation, we use four metrics introduced in  Sec.\,\ref{sec:general}: UA (unlearning accuracy) and MIA (membership inference attack-based unlearning efficacy) to assess unlearning performance, and RA (accuracy on the retain set) and TA (accuracy on the testing set) to evaluate the utility of the unlearned model. Also for image generation, we have additionally introduced the In-domain retain accuracy (IRA), Cross-domain retain accuracy (CRA), and FID metrics followed the setting of \textsc{UnlearnCanvas}\cite{zhang2024unlearncanvas}.



\noindent\textbf{BLO implementation for {\ours}.}
For image classification context, in the upper-level optimization, we use the proposed implicit gradient-based descent approach with setting the learning rate to $10^{-4}$ over 10 epochs. In the lower-level optimization, unlearning is performed with 3 epochs at the learning rate $10^{-2}$. The Hessian diagonalization parameter is set to $\lambda = 10^{-2}$. For image generation context, we set the learning rate to $10^{-4}$ over 6 epochs in the upper-level optimization, while setting the learning rate to $10^{-5}$ over 5 epochs in the lower-level optimization. The Hessian diagonalization parameter is set to $\lambda = 10^{-2}$. For watermark message selection, we use the same optimization approach as image classification context, while setting the learning rate to $\alpha = 10^{-3}$ over 20 epochs in the upper-level optimization. The Hessian diagonalization parameter is set to $\lambda = 10^{-3}$. All our experiments are conducted on one NVIDIA RTX A6000 48G GPU.


\subsection{Experiment results}

\noindent\textbf{Effectiveness of {\ours}-integrated unlearning.}
In \textbf{Tab.\,\ref{tab:cifar10_main}}, we present the performance of various MU methods, including the exact unlearning method Retrain and approximate unlearning methods (FT, GA, Sparse, and IU), \textit{with} and \textit{without} the integration of {\ours} on the (CIFAR-10, ResNet-18) setup. This integration is implemented under scenario $\mathcal{S_2}$--watermarked unlearning with unwatermarked evaluation--as defined in Sec.\,\ref{sec:general}, with the watermarking mechanism provided by {\ours}. Here our focus on unwatermarked evaluation allows us to examine whether watermarking can enhance MU performance even in evaluations on the original, unwatermarked forget set.
We provide several key observations from Tab.\,\ref{tab:cifar10_main} below.

\begin{table*}[htb]
    \centering 
    \caption{Performance of $\ours$ on additional datasets in \textsc{ResNet-18} (\textsc{CIFAR-10} and \textsc{SVHN} for random data forgetting, \textsc{ImageNet} for class-wise forgetting). The class-wise forgetting is conducted over 10\% of ImageNet classes. The other content format follows \cref{tab:cifar10_main}. 
    }
   \vspace*{-1em}
    \resizebox{0.95\linewidth}{!}{
    \begin{tabular}{c|ccc|ccc|ccc|ccc|ccc}
        \toprule[1pt]
        \midrule
         \multirow{2}{*}{Metric}&\multicolumn{3}{c|}{\textbf{Retrain}}&\multicolumn{3}{c|}{\textbf{GA}}&\multicolumn{3}{c|}{\textbf{FT}}&\multicolumn{3}{c|}{\textbf{Sparse}}&\multicolumn{3}{c}{\textbf{IU}}\\
         &\multicolumn{1}{c}{Original}&\multicolumn{1}{c}
         {$\ours$}&Diff&\multicolumn{1}{c}{Original}&\multicolumn{1}{c}{$\ours$}&Diff&\multicolumn{1}{c}{Original}&\multicolumn{1}{c}
         {$\ours$}&Diff&\multicolumn{1}{c}{Original}&\multicolumn{1}{c}
         {$\ours$}&Diff&\multicolumn{1}{c}{Original}&\multicolumn{1}{c}
         {$\ours$}&Diff\\
         \midrule
         \multicolumn{16}{c}{10\% random data forgetting, \textsc{CIFAR-100}}\\
         \midrule
         UA$\uparrow$&$24.88_{\pm0.13}$&$28.32_{\pm0.67}$&\textcolor{darkgreen}{3.44$\blacktriangle$}&$0.05_{\pm0.01}$&$2.29_{\pm0.12}$&\textcolor{darkgreen}{2.24$\blacktriangle$}&$0.98_{\pm0.06}$&$3.11_{\pm0.27}$&\textcolor{darkgreen}{2.13$\blacktriangle$}&$9.12_{\pm0.21}$&$13.31_{\pm0.11}$&\textcolor{darkgreen}{4.19$\blacktriangle$}&$3.19_{\pm0.09}$&$5.43_{\pm0.33}$&\textcolor{darkgreen}{2.24$\blacktriangle$}\\
         MIA$\uparrow$&$49.83_{\pm0.17}$&$49.96_{\pm0.25}$&\textcolor{darkgreen}{0.13$\blacktriangle$}&$2.40_{\pm0.21}$&$5.49_{\pm0.52}$&\textcolor{darkgreen}{3.09$\blacktriangle$}&$2.29_{\pm0.29}$&$5.89_{\pm0.32}$&\textcolor{darkgreen}{3.60$\blacktriangle$}&$12.56_{\pm0.72}$&$19.64_{\pm1.07}$&\textcolor{darkgreen}{7.08$\blacktriangle$}&$6.71_{\pm0.40}$&$10.78_{\pm0.33}$&\textcolor{darkgreen}{4.07$\blacktriangle$}\\
         RA$\uparrow$&$99.99_{\pm0.01}$&$99.38_{\pm0.21}$&\textcolor{blue}{0.61$\blacktriangledown$}&$99.97_{\pm0.02}$&$99.58_{\pm0.09}$&\textcolor{blue}{0.39$\blacktriangledown$}&$99.98_{\pm0.01}$&$99.91_{\pm0.03}$&\textcolor{blue}{0.07$\blacktriangledown$}&$96.32_{\pm0.07}$&$96.71_{\pm0.11}$&\textcolor{darkgreen}{0.39$\blacktriangle$}&$97.86_{\pm0.02}$&$97.11_{\pm0.03}$&\textcolor{blue}{0.75$\blacktriangledown$}\\

         TA$\uparrow$&$76.35_{\pm0.13}$&$73.31_{\pm0.22}$&\textcolor{blue}{3.04$\blacktriangledown
         $}&$75.91_{\pm0.19}$&$74.43_{\pm0.35}$&\textcolor{blue}{1.48$\blacktriangledown$}&$75.96_{\pm0.27}$&$74.83_{\pm0.14}$&\textcolor{blue}{1.13$\blacktriangledown
         $}&$70.43_{\pm0.49}$&$69.77_{\pm0.26}$&\textcolor{blue}{0.66$\blacktriangledown
         $}&$72.75_{\pm0.25}$&$72.53_{\pm0.16}$&\textcolor{blue}{0.22$\blacktriangledown
         $}\\
         \midrule
         \multicolumn{16}{c}{10\% random data forgetting, \textsc{SVHN}}\\
         \midrule
         UA$\uparrow$&$7.49_{\pm0.16}$&$8.79_{\pm0.29}$&\textcolor{darkgreen}{1.30$\blacktriangle$}&$0.02_{\pm0.01}$&$2.18_{\pm0.17}$&\textcolor{darkgreen}{2.16$\blacktriangle$}&$1.98_{\pm0.23}$&$4.31_{\pm0.20}$&\textcolor{darkgreen}{2.33$\blacktriangle$}&$5.01_{\pm0.17}$&$7.66_{\pm0.12}$&\textcolor{darkgreen}{2.65$\blacktriangle$}&$4.94_{\pm0.12}$&$7.02_{\pm0.47}$&\textcolor{darkgreen}{2.08$\blacktriangle$}\\
         MIA$\uparrow$&$17.78_{\pm0.14}$&$17.67_{\pm0.15}$&\textcolor{blue}{0.11$\blacktriangledown$}&$0.30_{\pm0.06}$&$3.67_{\pm0.22}$&\textcolor{darkgreen}{3.37$\blacktriangle$}&$5.35_{\pm0.19}$&$11.69_{\pm0.55}$&\textcolor{darkgreen}{6.34$\blacktriangle$}&$10.62_{\pm0.15}$&$13.81_{\pm0.38}$&\textcolor{darkgreen}{3.19$\blacktriangle$}&$7.23_{\pm0.09}$&$8.96_{\pm0.25}$&\textcolor{darkgreen}{1.73$\blacktriangle$}\\
         RA$\uparrow$&$100.00_{\pm0.00}$&$99.95_{\pm0.02}$&\textcolor{blue}{0.05$\blacktriangledown$}&$99.97_{\pm0.01}$&$99.00_{\pm0.13}$&\textcolor{blue}{0.97$\blacktriangledown$}&$98.98_{\pm0.07}$&$99.59_{\pm0.11}$&\textcolor{darkgreen}{0.61$\blacktriangle$}&$99.14_{\pm0.05}$&$98.57_{\pm0.12}$&\textcolor{blue}{0.57$\blacktriangledown$}&$99.46_{\pm0.02}$&$99.20_{\pm0.04}$&\textcolor{blue}{0.26$\blacktriangledown$}\\

         TA$\uparrow$&$93.71_{\pm0.42}$&$93.63_{\pm0.28}$&\textcolor{blue}{0.08$\blacktriangledown
         $}&$92.59_{\pm0.57}$&$92.51_{\pm0.18}$&\textcolor{blue}{0.08$\blacktriangledown$}&$92.89_{\pm0.13}$&$93.40_{\pm0.37}$&\textcolor{darkgreen}{0.51$\blacktriangle$}&$93.70_{\pm0.08}$&$93.22_{\pm0.17}$&\textcolor{blue}{0.48$\blacktriangledown$}&$91.13_{\pm0.15}$&$90.46_{\pm0.11}$&\textcolor{blue}{0.67$\blacktriangledown
         $}\\
         \midrule
         \multicolumn{16}{c}{Class-wise forgetting, \textsc{ImageNet}}\\
         \midrule
         UA$\uparrow$&$74.33_{\pm0.62}$&$76.45_{\pm0.49}$&\textcolor{darkgreen}{2.12$\blacktriangle$}&$61.10_{\pm0.45}$&$67.23_{\pm1.36}$&\textcolor{darkgreen}{6.13$\blacktriangle$}&$45.52_{\pm0.86}$&$53.71_{\pm0.59}$&\textcolor{darkgreen}{8.19$\blacktriangle$}&$80.25_{\pm0.59}$&$85.09_{\pm0.37}$&\textcolor{darkgreen}{4.84$\blacktriangle$}&$50.34_{\pm0.56}$&$53.71_{\pm0.86}$&\textcolor{darkgreen}{3.37$\blacktriangle$}\\
         MIA$\uparrow$&$98.30_{\pm0.17}$&$98.69_{\pm0.08}$&\textcolor{darkgreen}{0.39$\blacktriangle$}&$98.71_{\pm0.16}$&$99.64_{\pm0.22}$&\textcolor{darkgreen}{0.93$\blacktriangle$}&$96.35_{\pm0.34}$&$99.56_{\pm0.27}$&\textcolor{darkgreen}{3.21$\blacktriangle$}&$98.20_{\pm0.20}$&$99.94_{\pm0.33}$&\textcolor{darkgreen}{3.21$\blacktriangle$}&$95.78_{\pm0.18}$&$95.88_{\pm0.13}$&\textcolor{darkgreen}{0.10$\blacktriangle$}\\
         RA$\uparrow$&$65.80_{\pm0.31}$&$64.73_{\pm0.29}$&\textcolor{blue}{1.07$\blacktriangledown$}&$63.37_{\pm0.76}$&$63.05_{\pm0.52}$&\textcolor{blue}{0.32$\blacktriangledown$}&$65.86_{\pm0.46}$&$65.68_{\pm0.19}$&\textcolor{blue}{0.18$\blacktriangledown$}&$64.34_{\pm0.83}$&$62.41_{\pm0.08}$&\textcolor{blue}{1.93$\blacktriangledown$}&$66.63_{\pm0.36}$&$66.24_{\pm0.12}$&\textcolor{blue}{0.39$\blacktriangledown$}\\

         TA$\uparrow$&$65.48_{\pm0.63}$&$64.21_{\pm0.77}$&\textcolor{blue}{1.27$\blacktriangledown
         $}&$63.22_{\pm0.47}$&$63.10_{\pm0.40}$&\textcolor{blue}{0.12$\blacktriangledown$}&$64.89_{\pm0.18}$&$64.37_{\pm0.52}$&\textcolor{blue}{0.52$\blacktriangledown$}&$63.69_{\pm0.34}$&$63.22_{\pm0.23}$&\textcolor{blue}{0.47$\blacktriangledown$}&$65.80_{\pm0.24}$&$65.09_{\pm0.0.25}$&\textcolor{blue}{0.71$\blacktriangledown$}\\
         
         \midrule
         \bottomrule[1pt]
    \end{tabular}
    }
    \vspace*{-1.5em}
    \label{tab:extend_main}
\end{table*}

\textbf{First}, integrating {\ours} into the unlearning process enhances unlearning effectiveness across all evaluated methods, as evidenced by improved UA and MIA scores compared to their performance without {\ours} (\textit{i.e.}, `Original' in Tab.\,\ref{tab:cifar10_main}). As noted in Sec.\,\ref{sec:general}, UA is defined as $1 -$(accuracy on the forget set $\mathcal{D}_f$), while the MIA score reflects the accuracy of the MIA detector in correctly identifying forget samples as non-training data. Therefore, an increase in UA and MIA scores indicates more effective unlearning.
Another notable observation is that the use of {\ours} further enhances the unlearning effectiveness of {Retrain} for data forgetting, indicating the proactive unlearning capability provided by {\ours}.

\begin{table}[b]
    \centering 
   \vspace*{-2em}
    \caption{Performance of {\ours} with watermarked evaluation. We choose Retrain as the unlearning method and compare the resulting performance with unwatermarked evaluation under (\textsc{CIFAR-10}, \textsc{ResNet-18}).
    }
   \vspace*{-1em}
    \resizebox{0.65\linewidth}{!}{
    \begin{tabular}{c|ccc}
         \toprule[1pt]
         \midrule
         Metric&\multicolumn{1}{c}{Unwatermarked}&\multicolumn{1}{c}
         {Watermarked}&Diff\\
         \midrule
         UA$\uparrow$&$10.01_{\pm0.69}$&$15.84_{\pm1.32}$&\textcolor{darkgreen}{5.83$\blacktriangle$}\\
         MIA$\uparrow$&$19.33_{\pm0.05}$&$26.55_{\pm0.15}$&\textcolor{darkgreen}{7.22$\blacktriangle$}\\
         RA$\uparrow$&$99.93_{\pm0.01}$&$97.63_{\pm0.76}$&\textcolor{blue}{2.30$\blacktriangledown$}\\
         TA$\uparrow$&$90.63_{\pm0.02}$&$88.79_{\pm0.55}$&\textcolor{blue}{1.84$\blacktriangledown
         $}\\
         
         \midrule
         \bottomrule[1pt]
    \end{tabular}
    }
    \label{tab:wm_for_eval}
    \vspace*{-0.5em}
\end{table}

\textbf{Second}, we observe that {\ours}-integrated unlearning approaches result in a slight decrease in model utility, as indicated by RA and TA scores. However, the gains in unlearning effectiveness (evidenced by higher UA and MIA scores) outweigh this slight reduction in TA and RA. This likely reflects the inherent tradeoff between enhanced unlearning effectiveness and preserved model utility \cite{jia2023model}. 
\textbf{Third}, we observe that integrating {\ours} is resilient to the choice of MU method. Recall that {\ours} was designed using the GradDiff unlearning objective in \eqref{eq: lower_level_MU}, closely aligning with methods like GA and FT. However, integrating {\ours} with other MU methods, such as Sparse and IU, also   enhances unlearning effectiveness, highlighting its broad applicability.

Additional results demonstrating the effectiveness of {\ours} on other datasets, including \textsc{CIFAR-100}, \textsc{SVHN}, and \textsc{ImageNet}, are presented in \textbf{Tab.\,\ref{tab:extend_main}}, showing consistent performance trends as observed in Tab.\,\ref{tab:cifar10_main}.

\noindent\textbf{Effectiveness of {\ours} in \textit{watermarked} evaluation.}
In \textbf{Tab.\,\ref{tab:wm_for_eval}}, we extend the unlearning evaluation of {\ours} from  Tab.\,\ref{tab:cifar10_main} to the watermarked evaluation scenario, where the evaluation sets are also watermarked using {\ours}. As observed, test-time watermarking further enhances unlearning effectiveness (UA and MIA) while slightly reduces utility performance (RA and TA) compared to unwatermarked evaluation. This suggests that watermarking can also serve as an effective test-time strategy to boost MU performance for a given forget set.

\begin{table*}[htb]
\vspace*{-1em}
    \centering 
    \caption{Performance of $\ours$ on worst-case challenging forget sets. For data-wise scenarios, we conduct worst-case forgetting on \textsc{CIFAR-10} using \textsc{ResNet-18}. For class-wise scenarios, we conducted worst-case forgetting on \textsc{ImageNet}. The data-wise forgetting ratio is 10\% of the whole training dataset and the class-wise forgetting ratio is 10\% of the entire classes.
    }
    \vspace*{-1em}
    \resizebox{0.8\linewidth}{!}{
    \begin{tabular}{c|ccc|ccc|ccc}
         \toprule[1pt]
         \midrule
         \multirow{2}{*}{Metric}&\multicolumn{3}{c|}{\textbf{Retrain}}&\multicolumn{3}{c|}{\textbf{GA}}&\multicolumn{3}{c}{\textbf{FT}}\\
         &\multicolumn{1}{c}{Original}&\multicolumn{1}{c}
         {$\ours$}&Diff&\multicolumn{1}{c}{Original}&\multicolumn{1}{c}{$\ours$}&Diff&\multicolumn{1}{c}{Original}&\multicolumn{1}{c}{$\ours$}&Diff\\
         \midrule
         \multicolumn{10}{c}{Worst-case data-wise forgetting, \textsc{CIFAR-10}}\\
         \midrule
         UA$\uparrow$&$0.00_{\pm0.00}$&$2.11_{\pm0.14}$&\textcolor{darkgreen}{2.11$\blacktriangle$}&$0.01_{\pm0.00}$&$1.73_{\pm0.02}$&\textcolor{darkgreen}{1.72$\blacktriangle$}&$0.03_{\pm0.01}$&$2.05_{\pm0.06}$&\textcolor{darkgreen}{2.02$\blacktriangle$}\\
         MIA$\uparrow$&$0.00_{\pm0.00}$&$3.03_{\pm0.09}$&\textcolor{darkgreen}{3.03$\blacktriangle$}&$0.33_{\pm0.05}$&$2.78_{\pm0.13}$&\textcolor{darkgreen}{2.45$\blacktriangle$}&$0.71_{\pm0.06}$&$3.40_{\pm0.31}$&\textcolor{darkgreen}{2.69$\blacktriangle$}\\
         RA$\uparrow$&$100.00_{\pm0.00}$&$99.87_{\pm0.02}$&\textcolor{blue}{0.13$\blacktriangledown$}&$99.13_{\pm0.03}$&$98.87_{\pm0.05}$&\textcolor{blue}{0.26$\blacktriangledown$}&$99.66_{\pm0.02}$&$99.56_{\pm0.02}$&\textcolor{blue}{0.10$\blacktriangledown$}\\

         TA$\uparrow$&$94.78_{\pm0.11}$&$94.62_{\pm0.23}$&\textcolor{blue}{0.16$\blacktriangledown
         $}&$94.18_{\pm0.11}$&$94.38_{\pm0.26}$&\textcolor{darkgreen}{0.20$\blacktriangle$}&$94.47_{\pm0.17}$&$93.75_{\pm0.14}$&\textcolor{blue}{0.72$\blacktriangledown
         $}\\
         \midrule
         \multicolumn{10}{c}{Worst-case class-wise forgetting, \textsc{ImageNet}}\\
         \midrule
         UA$\uparrow$&$48.49_{\pm0.14}$&$51.32_{\pm0.33}$&\textcolor{darkgreen}{2.83$\blacktriangle$}&$42.49_{\pm0.52}$&$50.21_{\pm0.43}$&\textcolor{darkgreen}{7.72$\blacktriangle$}&$33.56_{\pm0.06}$&$40.10_{\pm0.15}$&\textcolor{darkgreen}{6.54$\blacktriangle$}\\
         MIA$\uparrow$&$98.63_{\pm0.12}$&$98.76_{\pm0.21}$&\textcolor{darkgreen}{0.13$\blacktriangle$}&$98.25_{\pm0.14}$&$99.47_{\pm0.17}$&\textcolor{darkgreen}{1.22$\blacktriangle$}&$96.56_{\pm0.26}$&$99.82_{\pm0.08}$&\textcolor{darkgreen}{3.26$\blacktriangle$}\\
         RA$\uparrow$&$65.62_{\pm0.41}$&$64.96_{\pm0.28}$&\textcolor{blue}{0.66$\blacktriangledown$}&$62.83_{\pm0.20}$&$62.54_{\pm0.44}$&\textcolor{blue}{0.29$\blacktriangledown$}&$66.85_{\pm0.35}$&$66.52_{\pm0.18}$&\textcolor{blue}{0.33$\blacktriangledown$}\\

         TA$\uparrow$&$65.59_{\pm0.14}$&$65.31_{\pm0.57}$&\textcolor{blue}{0.28$\blacktriangledown
         $}&$63.14_{\pm0.15}$&$63.20_{\pm0.29}$&\textcolor{blue}{0.06$\blacktriangledown$}&$65.73_{\pm0.28}$&$64.35_{\pm0.72}$&\textcolor{blue}{1.38$\blacktriangledown$}\\
         
         \midrule
         \bottomrule[1pt]
    \end{tabular}
    }
    \label{tab:worst-case}
    \vspace*{-1em}
\end{table*}

\begin{figure*}[htb]
    \centering
    \includegraphics[width=0.8\linewidth]{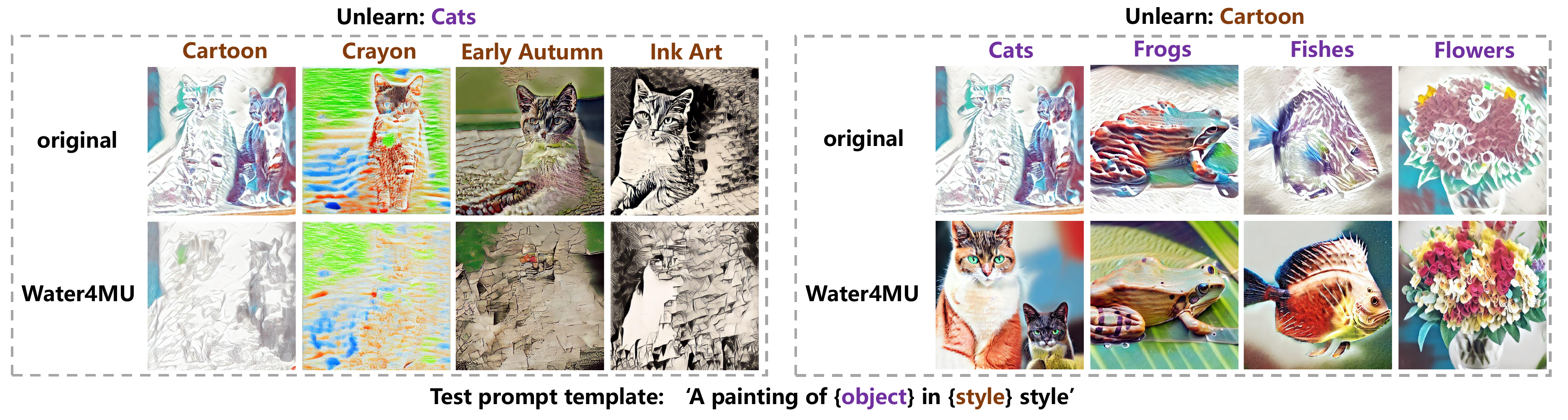}
    \vspace*{-1em}
    \caption{Examples of generated images by SD w/ and w/o MU with {\ours}. Concept types are distinguished by color: \textcolor{brown}{styles} in brown and \textcolor{Plum}{objects} in purple. The first row serves as a reference for comparison before unlearning.}
    \label{fig: unlearn_sd}
    \vspace*{-1.8em}
\end{figure*}

\noindent\textbf{Effectiveness of {\ours} on ``challenging forgets'' \cite{fan2025challenging}.}
In \textbf{Tab.\,\ref{tab:worst-case}}, we evaluate the performance of {\ours} on ``challenging forgets'', a worst-case forget set comprising the most difficult training samples or image classes to unlearn.  As we can see,  integrating {\ours} into Retrain, GA, and FT shows a clear enhancement in unlearning effectiveness, as evidenced by improved UA and MIA scores across all datasets. Comparing these results with random data forgetting on CIFAR-10 in Tab.\,\ref{tab:cifar10_main} and standard class-wise forgetting on ImageNet in Tab.\,\ref{tab:extend_main}, we observe that unlearning becomes notably more challenging in the worst-case scenario in Tab.\,\ref{tab:worst-case}. Nevertheless, {\ours} provides a promising solution to further improve existing MU methods, even in these challenging unlearning scenarios.

\begin{table}[b]
\vspace*{-2em}
    \centering 
    \caption{Performance overview of {\ours} evaluated on \textsc{UnlearnCanvas}. The performance metrics include UA (unlearning accuracy), IRA (in-domain retain accuracy), CRA (cross-domain retain accuracy), and FID. Results are averaged over all the style and object unlearning cases.
    }
    \vspace*{-1em}
    \resizebox{0.9\linewidth}{!}{
    \begin{tabular}{c|ccc|ccc|c}
         \toprule[1pt]
         \midrule
         \multirow{2}{*}{Metric}&\multicolumn{3}{c|}{\textbf{Style Unlearning}}&\multicolumn{3}{c|}{\textbf{Object Unlearning}}&\multicolumn{1}{c}{\textbf{}}\\
         &\multicolumn{1}{c}{UA$\uparrow$}&\multicolumn{1}{c}
         {IRA$\uparrow$}&CRA$\uparrow$&\multicolumn{1}{c}{UA$\uparrow$}&\multicolumn{1}{c}{IRA$\uparrow$}&CRA$\uparrow$&\multicolumn{1}{c}{FID$\downarrow$}\\
         \midrule
        ESD&98.58&80.97&93.96&92.15&55.78&44.23&65.55\\
        FMN&88.48&56.77&46.60&45.64&90.63&73.46&131.37\\
        UCE&98.40&60.22&47.71&94.31&39.35&34.67&182.01\\
        \midrule
        Ours&93.49&92.75&94.01&82.24&85.60&81.52&86.62\\
         \midrule
         \bottomrule[1pt]
    \end{tabular}
}
    \label{tab:sd_unlearn}
    \vspace*{-0.5em}
\end{table}

\noindent\textbf{An extended study: effectiveness of WATER4MU in image generation.} We further demonstrate the efficacy of our approach in prompt-wise forgetting for text-to-image generation. We use the Stable Diffusion (SD) model \cite{rombach2022high} on the \textsc{UnlearnCanvas} dataset \cite{zhang2024unlearncanvas}, a benchmark dataset designed to evaluate the unlearning of painting styles and objects. In UnlearnCanvas, a text prompt used as the condition of image generation is given by `A painting of [Object Name] in [Style Name] Style.'. We consider 20 objects and 50 styles for unlearning. \textbf{In Tab.\,\ref{tab:sd_unlearn}}, we provide an overview of the performance of {\ours}, using the evaluation metrics associated with UnlearnCanvas. \textbf{First}, as seen in Tab.\,\ref{tab:sd_unlearn}, we observe that {\ours} keeps high UA in both style and object unlearning (95\% and 82\%). \textbf{Second}, retainability of {\ours} is maintained at a high level. Both IRA and CRA exceeds 80\% in the context of style and object unlearning. Compared to other MU methods, {\ours} achieves the best balance between unlearning effectiveness and retainability. \textbf{Third}, {\ours} manages to maintain good generation quality (measured by FID). The above results indicates that {\ours} can be effectively adapted to prompt-wise unlearning tasks. \textbf{Fig.}\,\ref{fig: unlearn_sd} showcases the visualizations of the generations produced by unlearned SD with {\ours}. As observed, {\ours} achieves effective concept erasure in both style and object unlearning. 

\noindent\textbf{Ablation studies.} 
In Appx.\,\ref{app: exp_res}, we provide computational costs of {\ours}, a hyperparameter sensitivity analysis, the decoding performance of {\ours}, and the influence of different watermark message selection.

\section{Conclusion}
\label{sec:conclusion}

In this work, we present {\ours}, a novel machine unlearning (MU)-aware watermarking framework that introduces a data-centric approach to enhance unlearning effectiveness in image classification. By integrating digital watermarking into MU workflows, {\ours} complements existing model-centric unlearning methods, showcasing how watermarking can act as a powerful tool for controlled data erasure. Our findings demonstrate that {\ours} significantly improves unlearning effectiveness across diverse MU methods, including under ``challenging forgets'' scenarios.


\clearpage
\renewcommand{\bibname}{References}

{
    \small
       \bibliographystyle{unsrt}
    \bibliography{refs/MU,refs/MU_LLM_ref_SLiu, refs/peft}
}

\newpage
\twocolumn

\section*{\Large{Appendix}}
\setcounter{section}{0}
\setcounter{figure}{0}
\setcounter{table}{0}
\makeatletter 
\renewcommand{\thesection}{\Alph{section}}
\renewcommand{\theHsection}{\Alph{section}}
\renewcommand{\thefigure}{A\arabic{figure}} 
\renewcommand{\theHfigure}{A\arabic{figure}} 
\renewcommand{\thetable}{A\arabic{table}}
\renewcommand{\theHtable}{A\arabic{table}}
\makeatother

\renewcommand{\thetable}{A\arabic{table}}
\setcounter{equation}{0}
\renewcommand{\theequation}{A\arabic{equation}}

\appendix
\setcounter{page}{1} 
\section{BLO framework of $\ours$ for image generation}
\label{sec:sd}
In the context of prompt-wise forgetting in image generation, the BLO problem for {\ours} can be also cast as \eqref{eq: BLO_Water4MU}: 
\begin{align}
    \begin{array}{ll}
\displaystyle \minimize_{\boldsymbol{\psi}, \boldsymbol{\phi}}         &      \hat{\mathcal{L}} (\boldsymbol{\psi}, \boldsymbol{\phi},  \btheta_\mathrm{u}(\bpsi))  \\
     \text{subject to}    &  \btheta_\mathrm{u}(\bpsi)=\argmin_\btheta ~\mathcal{L}_{\mathrm{mu}} (\btheta; \hat{\mathcal D_\mathrm{f}}, \hat{\mathcal D_\mathrm{r}}),
    \end{array}
    \label{eq: BLO_Water4MU_sup}
\end{align}
In the lower-level optimization, we first obtain the watermarked dataset for unlearning. Then, we extend the use of Random Label (RL) to the image generation context for unlearning concept $c$. Also, to maintain the generation capability of the model, we introduce the regularization loss on the watermarked retain set. Finally, we can obtained an unlearned model $\btheta_\mathrm{u}$:
\begin{align}
 \btheta_\mathrm{u}(\bpsi)=\argmin_\btheta ~
\mathbb E_{\mathbf x \in \hat{\mathcal D_\mathrm{f}},t,\boldsymbol{\epsilon}}[||\boldsymbol{\epsilon}_\btheta(x_t|c')-\boldsymbol{\epsilon}_\btheta(x_t|c)||_2^2]\nonumber\\
 +\mathbb E_{\mathbf x \in \hat{\mathcal D_\mathrm{r}},t,\boldsymbol{\epsilon}}[||\boldsymbol{\epsilon}-\boldsymbol{\epsilon}_\btheta(x_t|c_r)||_2^2],  
\label{eq: lower_level_MU_sd}
\end{align}
where $c'\neq c$, $c_r$ is the prompt of $x \in \hat{\mathcal D_\mathrm{r}}$, $\hat{\mathcal D_\mathrm{f}}$ (or $\hat{\mathcal D_\mathrm{r}}$) denotes the watermarked forget (or retain) dataset, $\btheta$ is the pretrained generative model and $\boldsymbol{\beta}$ is a regularization parameter.

The design of the upper-level optimization follows \eqref{eq: upper_level_obj}:
\begin{align}
    \hat{\mathcal{L}} (\boldsymbol{\psi}, \boldsymbol{\phi},  \btheta_\mathrm{u}(\bpsi)) \Def & \underbrace{\mathcal{L}_{\mathrm{mu}} (\btheta_\mathrm{u}(\bpsi); \Df, \Dr)}_\text{(a) Unlearning validation} \nonumber \\
    & + \underbrace{
    \mathcal{L}_{\mathrm{wm}} (\boldsymbol{\psi}, \boldsymbol{\phi}; \mathbf m, \Df \cup \Dr)}_\text{(b) Watermarking validation},
    \label{eq: upper_level_obj_sd}
\end{align}
where $\ell_\mathrm{mu}$ is defined in \eqref{eq: lower_level_MU_sd} and $\mathcal{L}_{\mathrm{mu}} (\btheta_\mathrm{u}(\bpsi); \Df, \Dr)$ is to validate the lower-level unlearned model $\btheta_\mathrm{u}(\bpsi)$ on the unwatermarked dataset, and $\ell_\mathrm{wm}$ is the training loss of the watermarking network.
\section{Additional Experiment Setup}
\label{sec:setup}
\subsection{{\ours} for image classification}
For the exact unlearning method Retrain, the training process comprises 182 epochs, utilizing the SGD optimizer with a cosine-scheduled learning rate initially set to 0.1. For FT, the unlearning process takes 10 epochs, during which the optimal learning rate is searched within the range of [$10^{-3}$, $10^{-1}$]. For GA, the unlearning process spans 5 epochs with the interval [$10^{-5}$, $10^{-3}$]. Regarding the method Sparse, the unlearning-enabled model updating process also
takes 10 epochs, searching the optimal sparse ratio in the range [$10^{-6}$, $10^{-4}$] and exploring learning rate within [$10^{-3}$, $10^{-1}$]. Finally, for IU, the parameter $\alpha$ (associated with the WoodFisher Hessian Inverse approximation) is searched within the range [$1$, $20$]. 
\subsection{{\ours} for image generation}
We follow the settings in the \textsc{UnlearnCanvas} benchmark, selecting 20 objects and 50 styles for unlearning. The $\boldsymbol{\beta}$ is set at 0.5, with a batch size of 1. The sampling settings involve the use of DDPM, 100 time steps, and a conditional scale of 7.5.

\section{Additional Experiment Results}
\label{app: exp_res}
\subsection{Ablation study}
\label{sec:add}
\paragraph{The computational costs of $\ours$.}
We measure the run-time efficiency (RTE) of applying an MU method, \textit{i.e.}, its computation time. In Tab.\,\ref{tab:RTE}, we compared the RTE of different methods with and without using $\ours$ on (\textsc{CIFAR-10}, \textsc{ResNet-18}). As we can see, the introduction of $\ours$ does not hamper the computation efficiency, highlighting its practicality.
\begin{table}[h]
    \centering
    \caption{Performance of RTE (min) of different unlearning methods under unwatermarked forget/retain sets (Original) and $\ours$-induced watermarked forget/retain sets on (\textsc{CIFAR-10}, \textsc{ResNet-18}) for class-wise forgetting. }
    \vspace*{-1em}
       \resizebox{0.8\linewidth}{!}{
    \begin{tabular}{c|c|c|c|c|c}
    \toprule[1pt]
         \midrule
         MU&Retrain&GA&FT&Sparse&IU \\
         \midrule
         Original&42.35&0.25&2.50&2.52&3.25\\
         \midrule
         $\ours$&49.93&0.30&3.07&3.09&3.98\\
         \midrule
    \bottomrule[1pt]
    \end{tabular}
    }
    \label{tab:RTE}
\end{table}

\begin{table*}[htb]
    \centering 
    \caption{Performance of different unlearning methods under unwatermarked forget/retain sets (Orignal) and $\ours$-induced watermarked forget/retrain sets on (\textsc{CIFAR-100}, \textsc{ResNet-18}) and (\textsc{SVHN}, \textsc{ResNet-18}) for class-wise forgetting. 
    }
    \vspace*{-1em}
    \resizebox{0.7\linewidth}{!}{
    \begin{tabular}{c|ccc|ccc|ccc}
         \toprule[1pt]
         \midrule
         \multirow{2}{*}{Metric}&\multicolumn{3}{c|}{\textbf{Retrain}}&\multicolumn{3}{c|}{\textbf{GA}}&\multicolumn{3}{c}{\textbf{FT}}\\
         &\multicolumn{1}{c}{Original}&\multicolumn{1}{c}
         {$\ours$}&Diff&\multicolumn{1}{c}{Original}&\multicolumn{1}{c}{$\ours$}&Diff&\multicolumn{1}{c}{Original}&\multicolumn{1}{c}{$\ours$}&Diff\\
         \midrule
         \multicolumn{10}{c}{Class-wise forgetting, \textsc{ResNet-18}, \textsc{CIFAR-100}}\\
         \midrule
         UA&$100.00_{\pm0.00}$&$100.00_{\pm0.00}$&0.00-&$83.67_{\pm0.78}$&$90.46_{\pm0.43}$&\textcolor{darkgreen}{6.33$\blacktriangle$}&$24.73_{\pm2.63}$&$29.56_{\pm2.78}$&\textcolor{darkgreen}{4.83$\blacktriangle$}\\
         MIA&$100.00_{\pm0.00}$&$100.00_{\pm0.00}$&0.00-&$91.51_{\pm2.66}$&$98.22_{\pm0.12}$&\textcolor{darkgreen}{6.71$\blacktriangle$}&$45.61_{\pm0.3.31}$&$50.36_{\pm2.09}$&\textcolor{darkgreen}{4.75$\blacktriangle$}\\
         RA&$99.98_{\pm0.01}$&$98.87_{\pm0.05}$&\textcolor{blue}{1.11$\blacktriangledown$}&$91.56_{\pm0.54}$&$88.32_{\pm1.18}$&\textcolor{blue}{3.24$\blacktriangledown$}&$99.18_{\pm0.26}$&$99.03_{\pm0.72}$&\textcolor{blue}{0.15$\blacktriangledown$}\\

         TA&$74.43_{\pm0.23}$&$72.89_{\pm0.13}$&\textcolor{blue}{0.16$\blacktriangledown
         $}&$65.79_{\pm0.69}$&$64.61_{\pm0.19}$&\textcolor{blue}{1.18$\blacktriangledown$}&$74.33_{\pm0.14}$&$72.50_{\pm0.27}$&\textcolor{blue}{1.83$\blacktriangledown
         $}\\
         \midrule
         \multicolumn{10}{c}{Class-wise forgetting, \textsc{ResNet-18}, \textsc{SVHN}}\\
         \midrule
         UA&$100.00_{\pm0.00}$&$100.00_{\pm0.00}$&0.00-&$83.29_{\pm0.42}$&$89.07_{\pm0.13}$&\textcolor{darkgreen}{5.78$\blacktriangle$}&$16.98_{\pm4.60}$&$29.84_{\pm3.75}$&\textcolor{darkgreen}{12.86$\blacktriangle$}\\
         MIA&$100.00_{\pm0.00}$&$100.00_{\pm0.00}$&0.00-&$98.23_{\pm0.38}$&$99.61_{\pm0.09}$&\textcolor{darkgreen}{1.38$\blacktriangle$}&$85.57_{\pm2.32}$&$90.10_{\pm0.69}$&\textcolor{darkgreen}{4.53$\blacktriangle$}\\
         RA&$100.00_{\pm0.00}$&$99.96_{\pm0.02}$&\textcolor{blue}{0.04$\blacktriangledown$}&$99.51_{\pm0.21}$&$98.16_{\pm0.79}$&\textcolor{blue}{1.35$\blacktriangledown$}&$100.00_{\pm0.00}$&$98.64_{\pm0.56}$&\textcolor{blue}{1.36$\blacktriangledown$}\\

         TA&$95.82_{\pm0.04}$&$94.86_{\pm0.07}$&\textcolor{blue}{0.96$\blacktriangledown
         $}&$95.33_{\pm0.19}$&$94.16_{\pm0.08}$&\textcolor{blue}{1.17$\blacktriangledown$}&$95.95_{\pm0.18}$&$95.76_{\pm0.22}$&\textcolor{blue}{0.19$\blacktriangledown$}\\
         
         \midrule
         \bottomrule[1pt]
    \end{tabular}
    }
    \label{tab:add_class_appendix}
\end{table*}

\begin{figure}[htb]
    \centering
    \begin{tabular}{cc}
        \includegraphics[width=0.45\linewidth]{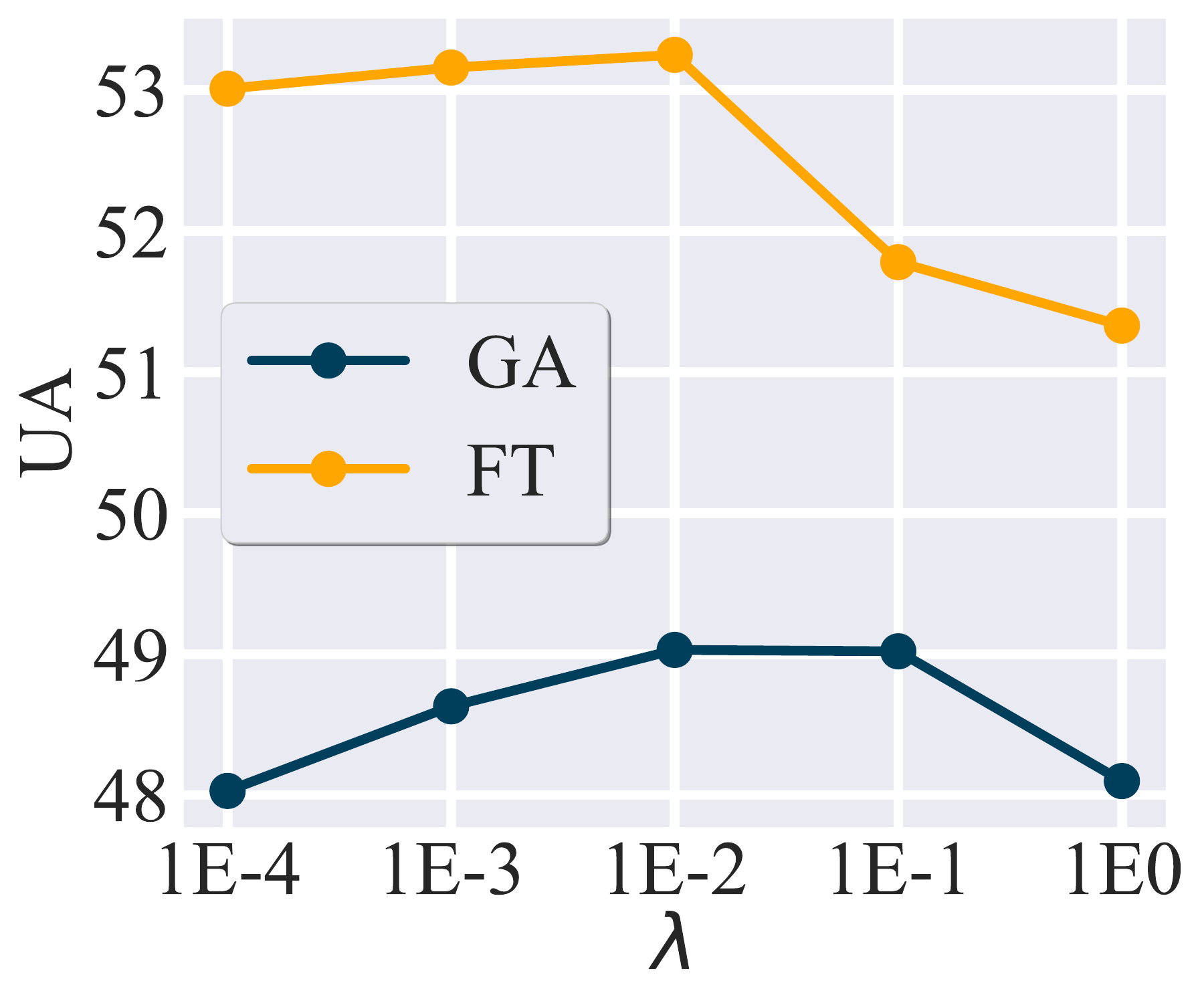} 
        & \includegraphics[width=0.45\linewidth]{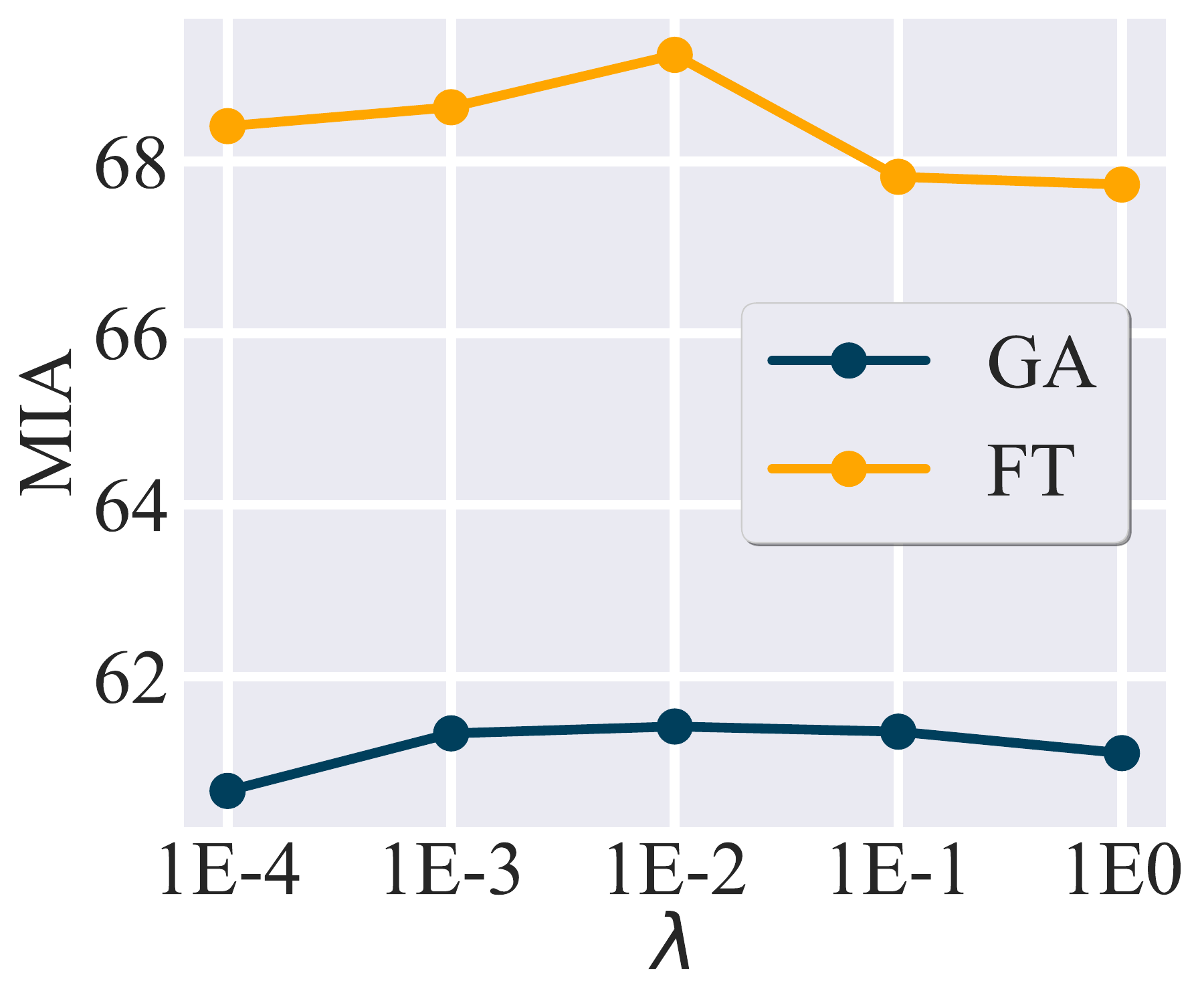}  \\ 
    \end{tabular}
    \vspace*{-5mm}
    \caption{Unlearning effectiveness (in terms of UA and MIA) of GA and FT against the choice of $\lambda$ in $\ours$ for class-wise forgetting under (\textsc{CIFAR-10}, \textsc{ResNet-18}). 
    }
    \vspace*{-1em}
    \label{fig:select_lambda}
\end{figure}

\begin{table*}[htb]
    \centering 
    \caption{Performance of different unlearning methods under unwatermarked forget/retain sets (Orignal) and $\ours$-induced watermarked forget/retrain sets on (\textsc{CIFAR-10}, \textsc{Swin Transformer}) and (\textsc{CIFAR-10}, \textsc{ResNet-50}) for class-wise forgetting. 
    }
    \vspace*{-1em}
    \resizebox{0.7\linewidth}{!}{
    \begin{tabular}{c|ccc|ccc|ccc}
         \toprule[1pt]
         \midrule
         \multirow{2}{*}{Metric}&\multicolumn{3}{c|}{\textbf{Retrain}}&\multicolumn{3}{c|}{\textbf{GA}}&\multicolumn{3}{c}{\textbf{FT}}\\
         &\multicolumn{1}{c}{Original}&\multicolumn{1}{c}
         {$\ours$}&Diff&\multicolumn{1}{c}{Original}&\multicolumn{1}{c}{$\ours$}&Diff&\multicolumn{1}{c}{Original}&\multicolumn{1}{c}{$\ours$}&Diff\\
         \midrule
         \multicolumn{10}{c}{Class-wise forgetting, \textsc{Swin Transformer}, \textsc{CIFAR-10}}\\
         \midrule
         UA&$100.00_{\pm0.00}$&$100.00_{\pm0.00}$&0.00-&$45.96_{\pm0.67}$&$55.43_{\pm0.50}$&\textcolor{darkgreen}{9.47$\blacktriangle$}&$94.89_{\pm1.56}$&$99.80_{\pm0.17}$&\textcolor{darkgreen}{4.91$\blacktriangle$}\\
         MIA&$100.00_{\pm0.00}$&$100.00_{\pm0.00}$&0.00-&$58.40_{\pm1.03}$&$64.82_{\pm1.89}$&\textcolor{darkgreen}{6.42$\blacktriangle$}&$97.86_{\pm1.21}$&$99.68_{\pm0.10}$&\textcolor{darkgreen}{1.82$\blacktriangle$}\\
         RA&$100.00_{\pm0.01}$&$99.71_{\pm0.07}$&\textcolor{blue}{0.29$\blacktriangledown$}&$99.85_{\pm0.26}$&$99.72_{\pm0.17}$&\textcolor{blue}{0.13$\blacktriangledown$}&$95.01_{\pm1.26}$&$93.45_{\pm0.92}$&\textcolor{blue}{1.56$\blacktriangledown$}\\
         TA&$85.63_{\pm2.10}$&$85.34_{\pm1.27}$&\textcolor{blue}{0.29$\blacktriangledown
         $}&$85.67_{\pm1.04}$&$85.32_{\pm1.41}$&\textcolor{blue}{0.35$\blacktriangledown$}&$80.58_{\pm2.16}$&$78.32_{\pm1.67}$&\textcolor{blue}{2.26$\blacktriangledown$}\\
         \midrule
         \multicolumn{10}{c}{Class-wise forgetting, \textsc{ResNet-50}, \textsc{CIFAR-10}}\\
         \midrule
         UA&$100.00_{\pm0.00}$&$100.00_{\pm0.00}$&0.00-&$36.76_{\pm1.28}$&$50.00_{\pm0.79}$&\textcolor{darkgreen}{13.24$\blacktriangle$}&$42.16_{\pm2.63}$&$59.83_{\pm2.11}$&\textcolor{darkgreen}{17.67$\blacktriangle$}\\
         MIA&$100.00_{\pm0.00}$&$100.00_{\pm0.00}$&0.00-&$59.60_{\pm0.65}$&$77.66_{\pm1.03}$&\textcolor{darkgreen}{18.06$\blacktriangle$}&$58.35_{\pm1.53}$&$67.75_{\pm1.93}$&\textcolor{darkgreen}{9.40$\blacktriangle$}\\
         RA&$100.00_{\pm0.00}$&$100.00_{\pm0.00}$&0.00-&$99.84_{\pm0.08}$&$99.53_{\pm0.19}$&\textcolor{blue}{0.31$\blacktriangledown$}&$98.76_{\pm0.14}$&$99.15_{\pm0.23}$&\textcolor{darkgreen}{0.39$\blacktriangle$}\\

         TA&$94.13_{\pm1.20}$&$94.02_{\pm0.98}$&\textcolor{blue}{0.11$\blacktriangledown
         $}&$93.67_{\pm0.72}$&$93.12_{\pm1.57}$&\textcolor{blue}{0.55$\blacktriangledown$}&$90.36_{\pm1.63}$&$90.96_{\pm1.49}$&\textcolor{darkgreen}{0.60$\blacktriangle$}\\
         
         \midrule
         \bottomrule[1pt]
    \end{tabular}
    }
    \label{tab:add_scale_appendix}
\end{table*}
\begin{table}[htb]
    \centering
    \caption{Number of nude body parts detected by Nudenet on I2P dataset with threshold 0.6.}
    \scriptsize
    \setlength\tabcolsep{3.5pt}
\vspace{-0.4em}
    \vspace{-0.8em}
    \begin{tabular}{cccccccc}
        \toprule[1pt]
        \midrule
        \textbf{Method} & $\textbf{Breast}$ & $\textbf{Genitalia}$ & $\textbf{Buttocks}$ & $\textbf{Feet}$ & $\textbf{Belly}$ & $\textbf{Armpits}$ & $\textbf{Total}$ \\
        \midrule
        \textbf{SD v1.4} & 229 & 31 & 44 & 42 & 171 & 129 & 646 \\
        \textbf{ESD} & \textbf{22} & \textbf{6} & 5 & 24 & 31 & 33 & 121 \\
        \textbf{FMN} & 172 & 17 & 12 & 56 & 116 & 42 & 415 \\
        \textbf{UCE} & 50 & 14 & 11 & 20 & 55 & 36 & 186\\
        \textbf{\ours} & 29 & 15 & \textbf{5} & \textbf{10} & \textbf{29} & \textbf{21} & \textbf{109}\\
        \midrule
        \bottomrule[1pt]
        \hline
    \end{tabular}
    \vspace{-2em}
    \label{tab:nude}
\end{table}
\paragraph{Choice of $\lambda$ in diagonalization approximation of the Hessian matrix.}
We next examine the hyperparameter  $\lambda$ in Hessian's diagonalization approximation in  \eqref{eq: IG_IF_Simple} for  {\ours}. In our experiments, the default setting is $\lambda = 10^{-2}$. We conduct a more detailed examination of $\lambda$ in Fig.\,\ref{fig:select_lambda}. We can observe that $\lambda=10^{-2}$ is a reasonable option, and higher or lower value of $\lambda$ would reduce the effectiveness of $\ours$.

\paragraph{Decoding performance of $\ours$.}
While {\ours} enhances the unlearning effectiveness, the decoding performance of {\ours} as a watermarking method should also be maintained. We then use BER (Bit Error Rate) to measure the performance of {\ours} to decode the watermark message $\mathbf m$. In our experiments, we set the message length $L$ to $10$ by default. We evaluate the decoding performance of   {\hidden} used by {\ours}. We find that its average BER is merely $2.78\times 10^{-8}$ compared to $1\times 10^{-8}$ using the standard {\hidden} w/o taking into account MU. This indicates that $\ours$ preserves the watermarking network's encoding-decoding capabilities.

\subsection{Additional class-wise forgetting results}
As an expansion of Tab.\,\ref{tab:cifar10_main}, Tab.\,\ref{tab:add_class_appendix} presents the performance of  class-wise forgetting with and without the integration of $\ours$ on the additional setups (\textsc{CIFAR-100},\textsc{ResNet-18}) and (\textsc{SVHN},\textsc{ResNet-18}). Also, Tab.\,\ref{tab:add_scale_appendix} shows the performance of class-wise forgetting on the setups (\textsc{CIFAR-10},\textsc{Swin Transformer}) and (\textsc{CIFAR-10},\textsc{ResNet-50}) to further validate the scalability of $\ours$. Notably, the {\ours}-integrated unlearning methods demonstrate enhanced unlearning performance across both datasets and models, as evidenced by improvements in UA and MIA metrics. This improvement effectively balances model utility preservation, as reflected in RA and TA scores. Importantly, the gains in unlearning performance outweigh the losses in utility. These observations are consistent with the findings reported in Tab.\,\ref{tab:cifar10_main}.

\subsection{Effect of Watermark Message Selection in {\ours}.}

\begin{figure}[htb]
    \centering
    \includegraphics[width=0.8\linewidth]{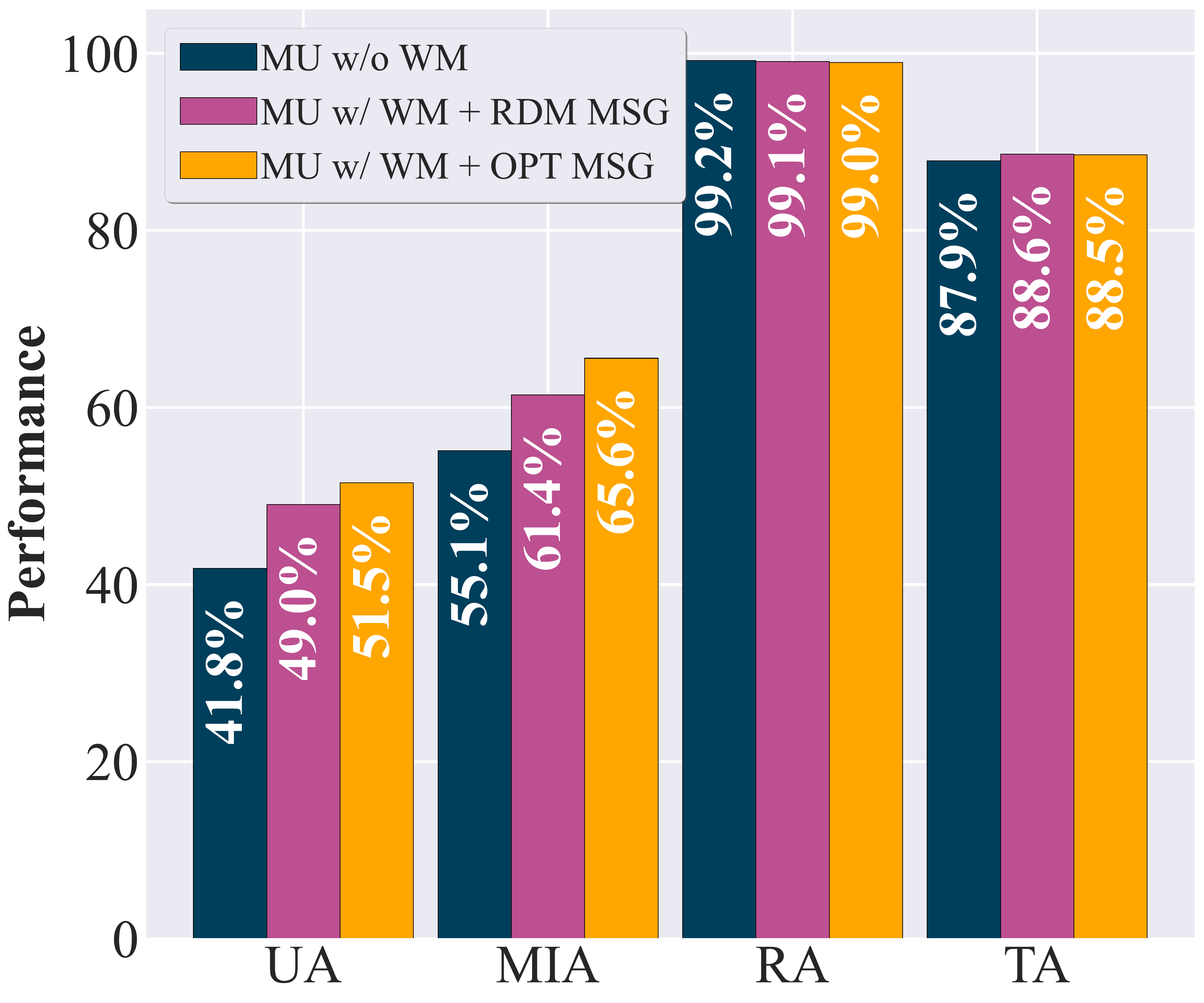}
    \vspace*{-1em}
    \caption{Performance of the optimized watermark message in $\ours$. We choose GA as the unlearning method and compare the unlearning performance among MU without $\ours$, MU with $\ours$ and MU with $\ours$ and optimized watermark message.
    }
    \label{fig: selection_of_m}
    \vspace*{-1em}
\end{figure}

In \textbf{Fig.\,\ref{fig: selection_of_m}}, we present the performance of the optimized watermark message, \textit{i.e.}, selecting the watermark message most favorable for unlearning, as described in Sec.\,\ref{sec:method}. As shown, using the optimized watermark message in {\ours} further enhances unlearning effectiveness (UA and MIA) without causing any degradation in model utility (TA and RA). This improvement is particularly notable when compared to the use of random watermark messages in {\ours}.

\begin{figure}[htb]
    \centering
    \includegraphics[width=0.8\linewidth]{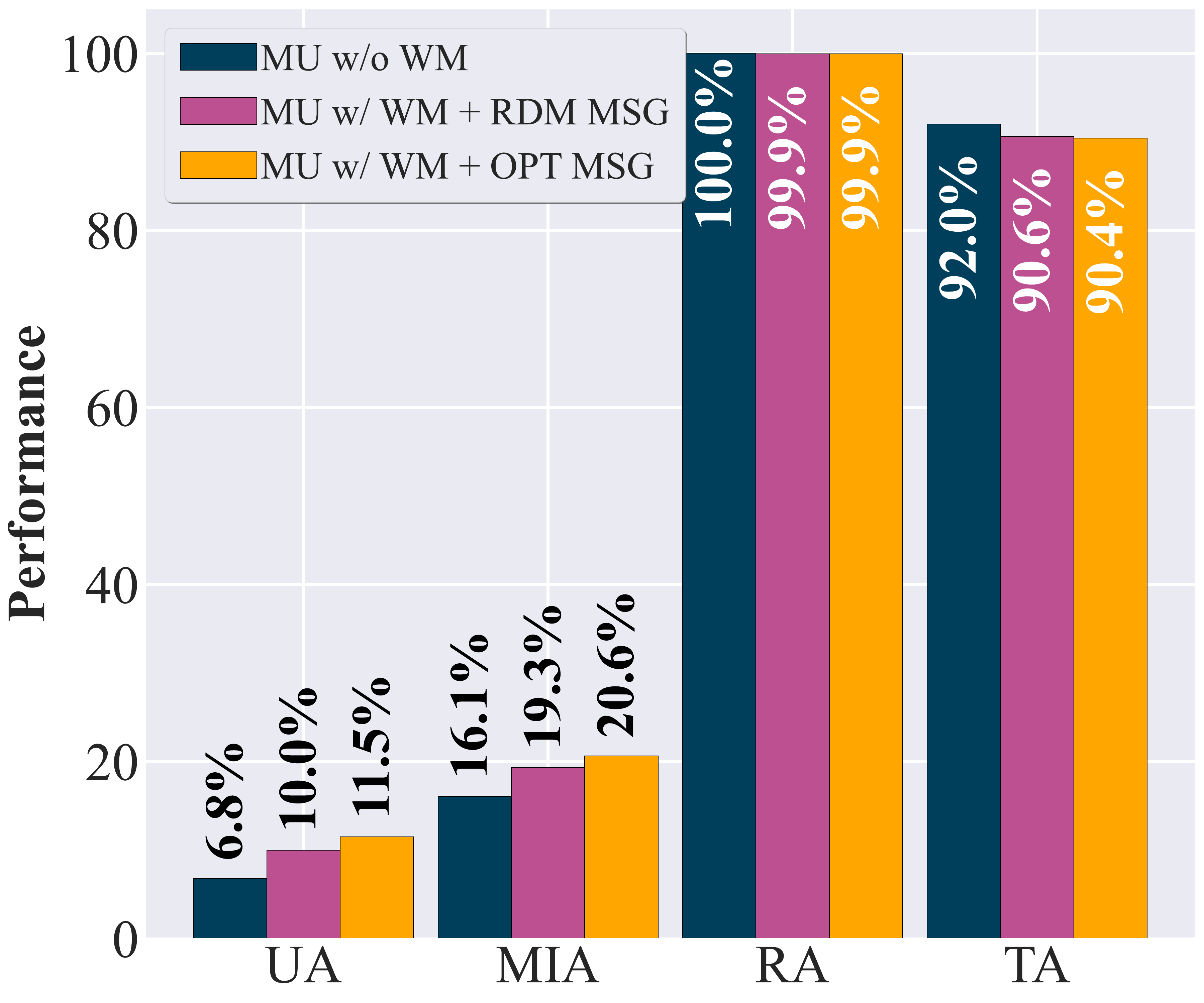}
    \vspace*{-1em}
    \caption{Performance of Retrain using the optimized watermark message in $\ours$ vs. baselines including MU without $\ours$ and MU with $\ours$ (but implemented using random watermark message).
    }
    \vspace*{-1em}
    \label{fig:select_watermark}
\end{figure}

We also provide additional results on the impact of watermark message selection on Retrain. Extended from Fig.\,\ref{fig: selection_of_m}, Fig.\,\ref{fig:select_watermark} presents the performance of the optimized watermark message on the exact unlearning method, Retrain. As we can see, using the optimized watermark message in $\ours$ also enhances unlearning effectiveness of Retrain without causing any degradation in model utility, when compared to the use of random watermark message in $\ours$. 

\begin{figure}[h]
    \centering
    \includegraphics[width=1\linewidth]{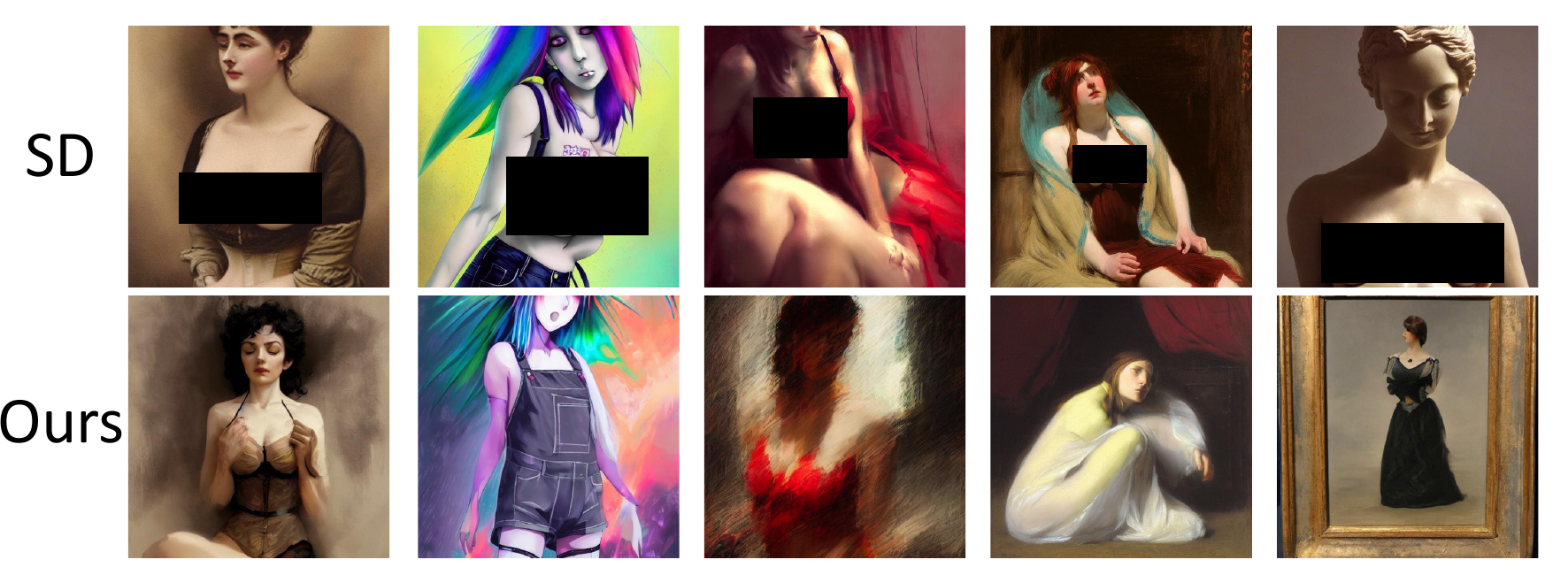}
    \vspace*{-1em}
    \caption{Qualitative results of nudity removing. All prompts originate from I2P dataset. The images in the top row are generated by SD, while the images in the bottom row are generated by MU with $\ours$.}
    \label{fig: unlearn_sd}
    \vspace*{-1em}
\end{figure}
\subsection{Evaluation on image generation safety tasks.}
We conduct experiments on the Inappropriate Image Prompts (I2P) dataset. Our evaluation focuses on the erasure of nudity. A total of 4703 images are generated with I2P prompts for each model. Then, Nudenet is introduced to detect nude body parts in these images. As shown in Tab.\,\ref{tab:nude}, we present the number of detected nude body parts across six categories. {\ours} achieves the best performance in preventing the generation of nude content, reducing the total instances from 646 to 109. We provide more examples on the erasure of the nudity concept in Fig.\,\ref{fig: unlearn_sd}.

\section{Limitations}
While {\ours} demonstrates promising unlearning effectiveness, the introduction of the watermarking network and the computation of higher-order derivatives in the BLO process will both incur additional training overhead. Further work is needed to assess the scalability of {\ours} to larger datasets and models, particularly in enhancing the efficiency of the bi-level optimization process used for watermarking design.

\end{document}